\documentclass[12pt]{article}
\usepackage{amssymb}
\usepackage{amsmath}
\usepackage{apacite}
\usepackage{latexsym}
\usepackage{epsfig}
\usepackage{graphicx}
\usepackage{subfigure}
\usepackage{wasysym}
\usepackage{threeparttable}
\usepackage{natbib}
\usepackage{color}
\usepackage{epstopdf}
\usepackage{bm}
\usepackage{float}
\usepackage{todonotes}
\usepackage{verbatim}

\usepackage{hyperref}

\definecolor{darkblue}{rgb}{0,0,0.5} 
\definecolor{darkred}{rgb}{0.5,0,0} 

\def\boxit#1{\vbox{\hrule\hbox{\vrule\kern6pt
          \vbox{\kern6pt#1\kern6pt}\kern6pt\vrule}\hrule}}

\def\bse{\begin{eqnarray*}}
\def\ese{\end{eqnarray*}}
\def\be{\begin{eqnarray}}
\def\ee{\end{eqnarray}}
\def\bq{\begin{equation}}
\def\eq{\end{equation}}
\def\bse{\begin{eqnarray*}}
\def\ese{\end{eqnarray*}}

\begin{document}

\thispagestyle{empty} 
\baselineskip=28pt

\begin{center}
{\LARGE{\bf Spatial Deep Convolutional Neural Networks}}

\end{center}

\baselineskip=12pt

\vskip 2mm
\begin{center}
Qi Wang\footnote{(\baselineskip=10pt to whom correspondence should be addressed)
Department of Statistics, University of California, Santa Cruz,
1156 High Street, Santa Cruz, CA 95064, qwang113@ucsc.edu}, Paul A. Parker\footnote{\baselineskip=10pt Department of Statistics, University of California, Santa Cruz,
1156 High Street, Santa Cruz, CA 95064, paulparker@ucsc.edu},
and Robert Lund\footnote{\baselineskip=10pt Department of Statistics, University of California, Santa Cruz,
1156 High Street, Santa Cruz, CA 95064, rolund@ucsc.edu}
\\
\end{center}
%
%
%
%
\vskip 4mm
\baselineskip=12pt 
\begin{center}
{\bf Abstract}
\end{center}

Spatial prediction problems often use Gaussian process models, which can be computationally burdensome in high dimensions. Specification of an appropriate covariance function for the model can be challenging when complex non-stationarities exist. Recent work has shown that pre-computed spatial basis functions and a feed-forward neural network can capture complex spatial dependence structures while remaining computationally efficient. This paper builds on this literature by tailoring spatial basis functions for use in convolutional neural networks. Through both simulated and real data, we demonstrate that this approach yields more accurate spatial predictions than existing methods. Uncertainty quantification is also considered.

\baselineskip=12pt
\par\vfill\noindent
\par\medskip\noindent
\clearpage\pagebreak\newpage \pagenumbering{arabic}
\baselineskip=24pt

\section{Introduction}
\label{sec:intro}

Gaussian processes are staple models for spatially dependent data. Gaussian processes have been applied to wind speeds \citep{cellura2008wind}, groundwater quality \citep{hooshmand2011application}, climatology \cite{zhang2023efficient}, and air pollution concentrations \citep{araki2015application}. A common assumption with Gaussian processes is stationarity and isotropy:  the mean is constant and the covariance between two observations only depends on the distance between the observations. However, stationarity and isotropy are often violated in practice, especially with multi-scale processes \citep{kirsner2020multi}.  In short, the development of spatial methods that incorporate nonstationarity is needed.

Another issue with Gaussian processes involves the inversion of large-dimensional covariance matrices that arise in forecasting and Gaussian likelihood estimation. Covariance matrix inversion has a complexity of $\textit{O}(n^3)$ for $n$ observations. Hence, for many large spatial data sets, Gaussian process computations become infeasible.  However, recent methods have been developed to reduce computations with minimal loss of precision, including nearest neighbor Gaussian processes methods (NNGP) \citep{finley2020spnngp}, fixed-rank Kriging \citep{cressie2008fixed}, and INLA-SPDE methods \citep{rue2009approximate,lindgren2011explicit}.

A variety of models have been proposed to handle spatial nonstationarity. For example, \cite{gramacy2008bayesian} partition the spatial domain into subspaces where stationarity reasonably holds for each subspace via Bayesian CART methods. \cite{kim2005analyzing} also use this tactic, but use Voronoi partitioning on the spatial domain. Another approach accommodates  nonstationarity by warping or expanding the spatial domain. For example, \cite{bornn2012modeling} use dimension expansion, constructing a stationary structure on an expanded space; \cite{schmidt2003bayesian} and \cite{sampson1992nonparametric} warp the spatial domain, assuming stationarity in the warped space. Furthering this line of work, \cite{zammit2022deep} use deep neural networks to estimate the warping function. \cite{higdon1998process} base their analysis on a discrete convolutional process having a spatially and temporally varying kernel; \cite{lemos2009spatio,lemos2012conditionally} and \cite{kirsner2020multi} further this line of research. Recently, \cite{kirsner2020multi} used a multi-scale spatial kernel convolution method that specifies local process properties at different resolutions.

As the popularity of neural networks grows, its ability to model complex nonlinear spatial functions has increased, especially in nonstationarity settings. As a baseline example, \cite{cracknell2014geological} feed spatial coordinates into neural networks to predict nonstationary spatial processes. Building on this, \cite{chen2020deepkriging} introduce the so-called DeepKriging approach, which feeds a set of spatial basis functions and coordinates into a neural network. This typically outperforms neural networks based on spatial coordinates alone \citep{chen2020deepkriging}. \cite{nag2023spatio} extend DeepKriging applications to spatial-temporal models by adding temporal basis functions to the spatial basis functions. There, a quantile loss function is used on a long short-term memory (LSTM) neural network to model time dependence. \cite{nag2023bivariate} develop bivariate DeepKriging for spatial processes with bivariate responses. \cite{zammit2023spatial} feed spatial RBFs into a Bayesian neural network (BNN) \citep{neal2012bayesian}, showing that spatial BNNs can adequately describe common spatial processes. \cite{zhan2023neural} use a generalized least squares loss with weights set via the nearest neighbor Gaussian processes' precision matrix. This is accomplished by adding graph convolution layers after a multi-layer perceptron. Finally, \cite{daw2023reds} use random Fourier expansions and apply extreme learning machine and ensembling operations to avoid training parameters and conduct uncertainty quantification. See \cite{wikle2023statistical} for a recent summary of statistical deep learning approaches for spatial and spatio-temporal data.

Motivated by \cite{chen2020deepkriging}, this paper shows how to model spatial dependence with convolutional neural networks. Similar to \cite{chen2020deepkriging}, basis functions are employed; however, we evaluate ours on a grid of knot locations.  This essentially treats the basis functions as images, allowing us to work with convolutional neural networks, a powerful tool for analyzing image data \citep{krizhevsky2012imagenet}. We use random dropout layers to quantify prediction uncertainty \citep{gal2016dropout}. Our spatial deep convolutional neural network --- dubbed a SDCNN --- is shown to achieve superior predictive performance in non-stationarity settings. Furthermore, by using neural networks that utilize stochastic gradient descent (rather than invert a spatial covariance matrix), our approach is extremely computationally efficient, capable of scaling to massive datasets. Our methods also make use of GPUs to further speed computations. 

The remainder of this paper proceeds as follows. In \hyperref[sec:bg]{Section 2}, relevant background material is introduced, including feed-forward and convolutional neural networks and their relation to the methods in \cite{chen2020deepkriging}. \hyperref[sec:methods]{Section 3} presents our proposed SDCNN. \hyperref[sec:simulation]{Section 4} offers a simulation study, showing the efficacy of the methods on synthetic data.  \hyperref[sec:application]{Section 5} demonstrates method performance on temperature and soil carbon data sets. \hyperref[sec:discussion]{Section 6} provides concluding remarks along with some future research directions.

\section{Background}
\label{sec:bg}

Deep learning applications have exploded in recent years, partially due to their ability to make highly accurate predictions of complex nonlinear processes. In this short section, widely used neural networks such as the feed-forward and convolutional neural networks are sketched. Since the parameters of a neural network are generally estimated by optimizing a loss function, gradient descent approaches are considered. A description of the DeepKriging approach of \cite{chen2020deepkriging}, one of our baseline models, is included.

\subsection{Feed-Forward Neural Networks}

Feed-forward neural networks (FNNs), which model non-linear structures in independent data, contain an input layer, one or more fully connected layers, and an output layer. Consider a single hidden layer FNN that models the univariate responses $\{ Y_i \}_{i=1}^N$. A length-$r$ covariate (input) vector $\boldsymbol{x}_i$ exists to help explain $Y_i$. The input layer first receives $\boldsymbol{x}_i$. Then, a fully connected layer computes an affine transformation of the $\boldsymbol{x}_i$ and activates it via the non-linear activation function $f_{h}(\cdot)$:
\begin{equation}
\label{eqn:fnnintro}
\boldsymbol{h}_i := 
f_{h}(\boldsymbol{W}^\prime \boldsymbol{x}_{i} + \boldsymbol{c}_{h}), 
\quad 1 \leq i \leq N,
\end{equation}
where $f_{h}(\cdot)$ is an activation function and $\boldsymbol{W}$ and $\boldsymbol{c}_{h}$ are affine transformation parameters. Clarifying dimensions, $\boldsymbol{W}$ is an $r \times n_h$ matrix and $\boldsymbol{c}_{h}$ is an $n_h$-dimensional vector. In the deep learning literature, the components of $\boldsymbol{W}$ are called hidden layer weights and $\boldsymbol{c}_{h}$ biases; $n_h$ denotes the number of hidden layer nodes. The length $n_h$ vector $\boldsymbol{h}_i$, called the hidden units, is the output of the hidden layer. While (\ref{eqn:fnnintro}) is written in vector form for brevity, $f_{h}(\cdot)$ is univariate and applied element-wise to the affinely transformed covariates. This layer is fully-connected, and it can be combined with other layers depending on problem goals.

Activation functions are crucial components of neural networks as they inject non-linear features into the modeling procedure. \cite{sharma2017activation} summarize frequently used activation functions. Two of the most commonly used are the rectified linear unit (ReLU) function $f(x) = \max(0,x)$ and the Sigmoid function $f(x) = (1+\exp(-x))^{-1}$.

The hidden layer in a single hidden-layer FNN is followed by an output layer, which utilizes another activation function $f_{o}(\cdot)$ to predict $Y_i$ (hats denote predictions):
\[
\hat{Y}_i := 
f_{\rm o}(\boldsymbol{\xi}^\prime \boldsymbol{h}_i + c_{\rm o}), 
\quad 1 \leq i \leq N.
\]
The length $n_h$ vector $\boldsymbol{\xi}$ contains the output layer weights and $c_{\rm o}$ is a scalar intercept parameter. Here, $f_{\rm o}(\cdot)$ is a univariate output layer activation function, typically chosen based on the support set of the $Y_i$s.  For example, the identity activation function $f_{\rm o}(x) = x$ is often used with a continuous numeric response, while the Sigmoid function is often employed for binary $Y_i$.  The quantities $\boldsymbol{W}$, $\boldsymbol{c}_{h}$, $\boldsymbol{\xi}$, and $c_{\rm o}$ above are often chosen to optimize some loss function. 

With continuous $\{ Y_i \}$, the quadratic loss $\sum_{i=1}^N (Y_i-\hat{Y}_i)^2$ is typically minimized. As $\hat{Y}_i$ is a non-linear function of $\boldsymbol{x}_i$, optimal parameter estimates do not usually have a closed form. Nonetheless, optimization algorithms often efficiently numerically minimize the loss via gradient descent. Even in complicated neural networks, back-propagation can efficiently calculate the gradients involved in the numerical minimization \citep{rumelhart1986learning}.

\subsection{Convolutional Neural Networks}
Some geographical or image data have spatial structures that FNNs may not efficiently model.  An alternative, a convolutional neural network (CNN), is known to be a superior image analysis tool \citep{albawi2017understanding}.  CNNs stem from \cite{lecun1995convolutional} and more capably explore localized spatial structures.  In addition to having convolutional layers, a CNN may also have fully connected and/or pooling layers, depending on needs.   Convolutional layers apply a discrete convolution to the input image (regarded as matrix data) with a set of trained filters.  The output of each filter is a ``convolved image" that contains the Hadamard product of the input image and filter for each compatible submatrix. 

Let $\boldsymbol{\mathcal{K}}$ be the image's kernel function and $\boldsymbol{\mathcal{I}}$ the image to be convolved, both bivariate.  The discrete convolution operation $*$ obeys
\[
\boldsymbol{\mathcal{K}}(x,y) * \boldsymbol{\mathcal{I}}(x,y) = \sum_{i=1}^\infty \sum_{j=1}^\infty \boldsymbol{\mathcal{K}}(i,j)\boldsymbol{\mathcal{I}}(x-i,y-j).
\]
The summation is truncated so that the indices $x-i$ and $y-j$ lie in the observed range of the image.

Convolutions can illuminate certain patterns or features in the image, including edges, corners, or textures, depending on filter choices \citep{o2015introduction}. Outputs from convolutional layers can be vectorized and combined with the fully-connected layer to increase model flexibility. 

\subsection{DeepKriging}

Spatial processes are often modeled by Gaussian dynamics. Let $\{ Z(\boldsymbol{s}) \}$ be a univariate Gaussian spatial process over $\boldsymbol{s} \in \mathcal{R}^d$ and let $\boldsymbol{Z} = \{ Z(\boldsymbol{s}_1), \ldots, Z(\boldsymbol{s}_N) \}^T$ denote the Gaussian field recorded over the $N$ locations $\boldsymbol{s}_1, \ldots, \boldsymbol{s}_N$. The observation $Z(\boldsymbol{s}_i)$ is usually a noisy measurement from some latent process $\{ Y(\boldsymbol{s}) \}$:
\[
Z(\boldsymbol{s}_i) = Y(\boldsymbol{s}_i) + \varepsilon(\boldsymbol{s}_i),
\]
where $\{ {\varepsilon}(\boldsymbol{s}) \}$ is a Gaussian error process that is independent of $\{ Y(\boldsymbol{s}) \}$ and independent from site to site.  Here, $\{ Y(\boldsymbol{s}) \}$ is the latent Gaussian process of interest, which is often further decomposed into
\[
Y(\boldsymbol{s}_i) = \boldsymbol{x}(\boldsymbol{s}_i)^T \boldsymbol{\beta} + \nu(\boldsymbol{s}_i),
\]
where $\boldsymbol{x}(\boldsymbol{s}_i)$ is a length-$p$ vector of explanatory covariates at location $\boldsymbol{s}_i$ and $\boldsymbol{\beta}$ is a vector of unknown regression coefficients. Spatial dependence is incorporated by allowing $\{ \nu(\boldsymbol{s}) \}$ to be a zero mean Gaussian process with the unknown spatial covariance $\mbox{Cov}(\nu(\boldsymbol{s}_i), \nu(\boldsymbol{s}_j)):= C(\boldsymbol{s}_i, \boldsymbol{s}_j)$.  

Methods exist that model Gaussian processes through basis function expansions.  The Karhunen-Lo\`eve expansion is a common expansion that represents a random field $\{ \omega(\boldsymbol{s}) \}$ with covariance function $C(\cdot,\cdot)$ as
\[
\omega(\boldsymbol{s})= 
\sum_{i=1}^\infty \alpha_i \phi_i(\boldsymbol{s}),
\]
where $\{ \phi_i(\cdot)\}_{i=1}^\infty$ are orthogonal functions and $\{ \alpha_i \}_{i=1}^\infty$ are uncorrelated random variables. From this, a number of basis functions, not necessarily orthogonal, can be defined. For example, one can express $\omega(\boldsymbol{s}) = \boldsymbol{b}(\boldsymbol{s})^\prime \boldsymbol{\eta}$, where $\boldsymbol{b}(\boldsymbol{s})$ is a vector of spatial basis functions and $\boldsymbol{\eta}$ is a multivariate Gaussian distribution having a diagonal covariance matrix. An Eigen-decomposition of $\boldsymbol{b}(\cdot)$ effectively rewrites $\omega(\boldsymbol{s})$ with a Karhunen-Lo\`eve expansion.

Two applications of basis function expansions are Gaussian predictive processes \citep{banerjee2008gaussian} and fixed-rank Kriging \citep{cressie2008fixed}. Both have a discrete convolution representation \citep{higdon1998process, lemos2009spatio, lemos2012conditionally}.  Radial basis functions (RBFs) are used in \cite{chen2020deepkriging} to represent the spatial process, enabling them to make predictions at unobserved locations with FNNs.  There, grids of basis functions over several resolutions are generated, governed by three parameters: location, range, and kernel function. The multi-resolution model of \cite{nychka2015multiresolution}, which explores different spatial structure scales, is used. The basis functions are
\begin{gather}
\label{eqn:nychbasis}
\phi(r) = 
\begin{cases}
(1-r)^6(35r^2+18r+3)/3, & r \in [0,1] \\
0  &  \mbox{otherwise},
\end{cases},
\end{gather}
where $r$ is the Euclidean distance between an observation and a basis function knot location, scaled by the range parameter $\theta$. As such, for each observation $i$, the $J$ basis functions defined by
\[
\psi_{i,j} := \phi \left( \frac{||\boldsymbol{s}_i - \boldsymbol{c}_j||_2}{\theta} \right), \quad 1 \leq j \leq J, 
\]
are introduced.  Here, the Euclidean distance is used, $\boldsymbol{s}_i$ is the location of $i$th observation, and $\boldsymbol{c}_j$ is the location extracted from the $j$th basis function.

Afterward, \cite{chen2020deepkriging} input the scaled spatial coordinates and RBFs into a deep neural network. Parameters in the neural networks are estimated by minimizing a quadratic loss, and further predictions are subsequently made. This method has a smaller prediction mean squared error (compared to classical Kriging and FNNs with spatial coordinate input) and avoids covariance matrix inversion. 

\section{Methodology}
\label{sec:methods}

This section introduces our spatial deep convolutional neural network (SDCNN). While we still also use RBFs, they are employed in non-classical ways. For each resolution, the RBFs are stored in an image format; in this way, each observation has multiple ``image covariates" (one for each resolution). For the RBFs at each resolution, convolutional layers are used to first explore image properties.  Then, the image is reshaped into a long vector (flattened) and fed into fully connected layers.  

Our SDCNN also employs a random dropout layer before each weight layer, avoiding overfitting and producing prediction uncertainty estimates. \cite{gal2016dropout} interpret dropout layers probabilistically, enabling deep neural networks to quantify prediction uncertainty. A fully connected layer is built alongside the other covariates. Finally, we concatenate outputs from all previous layers and make predictions in the output layer.  More detail follows.

\subsection{Basis Function Generation}

To set up the neural network, spatial RBFs are first  generated as input covariates.  The R package \texttt{FRK}, created by \cite{zammit2023introduction}, accomplishes this and is used henceforth. In this package, the function \texttt{auto\_basis} allows users to select resolution parameters and kernel function type (Gaussian, bisquare, etc.). Elaborating, each basis function has a location and scale parameter that standardizes the distance between each observation and the location of the basis function knot. Also needed is a kernel function, which is chosen as Gaussian here. The basis function evaluated at the $i$th observation and $j$th knot, denoted by $\phi(\boldsymbol{s}_i, \boldsymbol{c}_j)$, is
\[
\phi(\boldsymbol{s}_i, \boldsymbol{c}_j) = \exp\left( -\frac{|| \boldsymbol{s}_i - \boldsymbol{c}_j ||_2}{2\sigma^2} \right),
\]
where $\boldsymbol{s}_i$ denotes the $i$th observation's location, $\boldsymbol{c}_j$ the $j$th basis function's knot location, and $\sigma$ is a scale parameter.  In this way, both spatial coordinates and RBFs inject spatial information about the observation at $\boldsymbol{s}_i$. 

Our RBFs are generated under several different resolutions, which enables them to capture spatial effects occurring at different scales (this is a multi-resolution model \citep{nychka2015multiresolution}). For the first resolution, the number of RBFs is relatively small, but the scale parameter is large, serving to explore coarse spatial structure over larger scales. For smaller scale behavior, subsequent RBF resolutions employ an increasing number of RBFs and a decreasing scale parameter, effectively allowing more localized spatial structures to be explored.

\begin{figure}[h]
\centering
\includegraphics[width=0.9\textwidth]{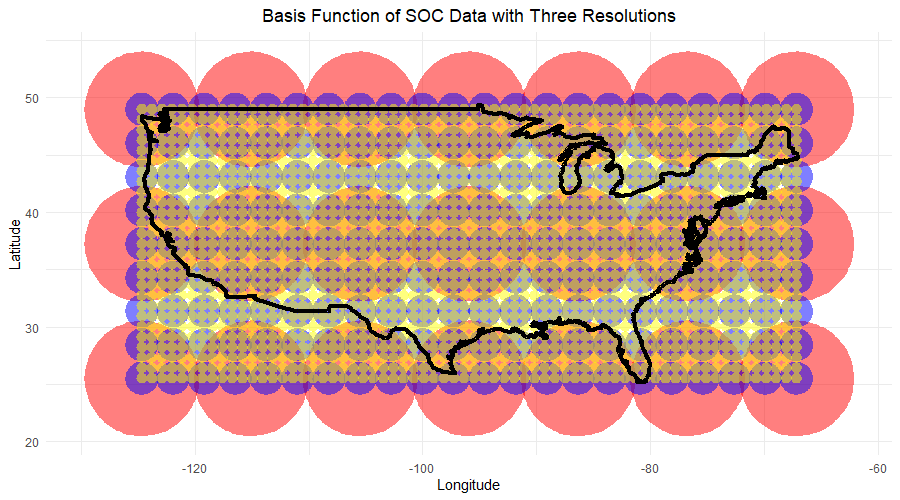}
\caption{Illustration of three RBF resolutions. The first resolution (red circles) has the largest scale and explores large-scale spatial effects. The second resolution (blue circles) offers a smaller more localized scale, while the third resolution (yellow circles) explores the finest localized spatial structure.}
\label{fig:basis_intro}
\end{figure}

An example with three RBF resolutions for United States soil carbon data, which is analyzed later, is depicted in Figure \ref{fig:basis_intro}.  Here, each RBF resolution has a natural grid structure. For the $k$th resolution in a two-dimensional spatial domain, $n_k \times m_k$ RBF factors for each observation arise; here, $n_k$ and $m_k$ denote the number of rows and columns, respectively, of the RBF's $k$th resolution. Given this structure, we treat the RBF values as an image (matrix) to fully take advantage of the power of CNNs.

\subsection{Uncertainty Quantification}

While neural networks are powerful tools capable of accurately predicting complex processes, the gradient based optimization methods used to fit them only provide point estimates of parameters and corresponding predictions. To explore estimation variability, we now delve into uncertainty quantification.

Dropout layers, introduced in \cite{srivastava2014dropout}, are layers that act to prevent overfitting while combining many potential models together. The term dropout refers to temporarily dropping a neural network unit and its subsequent influence. Our dropout layers specify a dropout probability $p$, and every column of the weights $\boldsymbol{W}$ in (\ref{eqn:fnnintro}) have probability $p$ of being set to zero, independently across components. This is quantified by setting 
\[
\Tilde{\boldsymbol{W}} := \mbox{diag}(\boldsymbol{\psi}) \boldsymbol{W},
\]
where $\boldsymbol{\psi} = (\psi_1, \psi_2, \ldots, \psi_{n_h})^\prime$ is a length-$n_h$ vector whose entries are only zero and one: $\psi_i \stackrel{iid} {\sim} \mbox{Bernoulli}(p)$ (iid means independent and identically distributed).  The dropout probability $p$ is usually chosen via cross-validation. In our work, $\bm{\psi}$ is resampled at every gradient descent iteration. Dropouts are typically done during training (i.e., gradient descent), but not when predicting. More recently, \cite{gal2016dropout} mathematically show that a dropout layer used before each hidden layer probabilistically approximates the scenario with deep Gaussian processes.  These authors further demonstrate how applying dropout layers for an ensemble of predictions can approximate model uncertainty, while simultaneously maintaining computational efficiency and accuracy. 

\subsection{SDCNNs}

After generating the RBFs, we feed each resolution's image into a convolutional layer. In our work, three different RBF resolutions are fed into three separate convolutional layers. The results from these convolutional layers are subsequently vectorized and concatenated with the output from the convolutional layer. These are then treated as input to a feed-forward layer. It is also possible to include additional input covariates within the feed-forward layer for spatial coordinates.

Taking the first resolution as an example, suppose that for each $\boldsymbol{s}_i$, the $m_1 \times n_1$ RBF matrix $\boldsymbol{B}_{i,1}$ is calculated, where ${(\boldsymbol{B}_{i,1})}_{m,n}$ is the RBF value associated with location $\boldsymbol{s}_i$ for the first resolution on the $m$th row and $n$th column. We then feed $\boldsymbol{B}_{i,1}$ into a convolutional layer containing $n_F$ filters, denoted by $\{ \boldsymbol{\mathcal{F}}_\ell\} _{\ell=1}^{n_F}$, where $\boldsymbol{\mathcal{F}}_\ell$ is a $2 \times 2$ matrix for all $\ell$.  Our work takes $n_F=128$ as is classic in the literature, but other settings can be tried.  Since the terms ``filter" and ``kernel" describe the same things in neural networks, we use the term filter henceforth to avoid confusion with the kernel terminology common in spatial statistics. The output from the convolutional layer for the $\ell$th filter, denoted by $\boldsymbol{C}_{i,\ell}$, is an $(n_1-1) \times (m_1-1)$ matrix with entries 
\[
(\boldsymbol{C}_{i,\ell})_{p,q} = \sum_{m=1}^2\sum_{n=1}^2
(\boldsymbol{B}_{i,1})_{p+m-1,q+n-1}
(\boldsymbol{\mathcal{F}}_\ell)_{m,n}.
\]

After computing the output from the convolutional layer, $\boldsymbol{C}_{i,\ell}$ is vectorized into a length-$(n_1-1)(m_1-1)$ vector. As there are $n_F$ similar filters, $n_F$ vectors with the same length are obtained after convolution and vectorization. Next, these vectors are stacked into one long vector, denoted by $\boldsymbol{\mathcal{C}}_{i,1}$, of length-$n_F(n_1-1)(m_1-1)$. 

The next step feeds $\boldsymbol{\mathcal{C}}_{i,1}$ into three feed-forward layers, with each layer having $N_h=100$ hidden nodes (this setting can also be changed) and using the same ReLU activation function $f$:
\begin{gather}
\begin{aligned}
     \label{eqn:sdcnnhidden}
    (\boldsymbol{\mathcal{H}}_i)_{1,1} &= f(\boldsymbol{W}_{1,1}^\prime \boldsymbol{\mathcal{C}}_{i,1} + \boldsymbol{c}_{1,1}), \\
        (\boldsymbol{\mathcal{H}}_i)_{1,2} &= f(\boldsymbol{W}_{1,2}^\prime (\boldsymbol{\mathcal{H}}_i)_{1,1} + \boldsymbol{c}_{1,2}), \\
        (\boldsymbol{\mathcal{H}}_i)_{1,3} &= f(\boldsymbol{W}_{1,3}^\prime (\boldsymbol{\mathcal{H}}_i)_{1,2} + \boldsymbol{c}_{1,3}),
\end{aligned}
\end{gather}
where $\boldsymbol{W}_{1,1}$, $\boldsymbol{W}_{1,2}$, and $\boldsymbol{W}_{1,3}$ are weight matrices of dimension $n_F(n_1-1)(m_1-1) \times N_h$ for $\boldsymbol{W}_{1,1}$ and $N_h \times N_h$ for the other two $\boldsymbol{W}$s.  Here, $\boldsymbol{c}_{1,1}$, $\boldsymbol{c}_{1,2}$, and $\boldsymbol{c}_{1,3}$ are length $N_h$ bias/intercept parameters. The vectors $(\boldsymbol{\mathcal{H}}_i)_{1,1}$, $(\boldsymbol{\mathcal{H}}_i)_{1,2}$, and $(\boldsymbol{\mathcal{H}}_i)_{1,3}$ each have length $N_h$.

Following similar steps, transformed RBFs for the other two resolutions, denoted by $(\boldsymbol{\mathcal{H}}_i)_{2,3}$ and $(\boldsymbol{\mathcal{H}}_i)_{3,3}$, are computed. The effects of any covariates $\boldsymbol{x}_i$ at location $\bm{s}_i$ can be quantified by feeding these into another feed-forward layer, again with $N_h$ nodes; this results in $(\boldsymbol{\mathcal{H}}_i)_{\boldsymbol{x}}$. At this point, the quantities $(\boldsymbol{\mathcal{H}}_i)_{1,3}$, $(\boldsymbol{\mathcal{H}}_i)_{2,3}$, $(\boldsymbol{\mathcal{H}}_i)_{3,3}$, and $(\boldsymbol{\mathcal{H}}_i)_{\boldsymbol{x}}$ are stacked into a length $4N_h$ vector denoted by $\boldsymbol{H}_i$. 

In our final step, $\boldsymbol{H}_i$ is fed into an output layer:
\[
O_i = \boldsymbol{W}_{\rm out}^\prime \boldsymbol{H}_i + c_{\rm out},
\]
where $\boldsymbol{W}_{\rm out}$ is a $4 N_h \times 1$ weight matrix and $c_{\rm out}$ is a constant (the identity activation function is used here). Here, $O_i$ serves as a prediction of $Y_i$ given the set of neural network parameters.  With stochastic gradient descent, one can numerically estimate all weight matrices and bias parameters by optimizing an appropriate loss function. In our case, a quadratic sum of squares loss is used with continuous responses. Random dropout layers are added before each hidden layer to quantify prediction uncertainty and avoid overfitting.

\section{Model Comparisons and Performance Evaluation Methods}
\label{sec:evaluations}
This section discusses how to compare our SDCNN to three other techniques.  Our first comparitative technique uses a Gaussian process fitted by integrated nested Laplace approximations (INLA).  The second and third comparitative techniques are neural networks with various inputs.  Several scoring rules and performance metrics are introduced below to judge the model fits.

\subsection{Competing Models}
We compare our SDCNN with the DeepKriging approach of \cite{chen2020deepkriging}, a baseline DNN with only spatial coordinates inputted, and a Gaussian process fitted via INLA. Comparisons are done via five-folded cross-validation (out of sample prediction) for one simulated data set and two real data sets. 

All code used in this paper has been uploaded to \url{https://github.com/qwang-77/Convolutional-Kriging} and the models are fitted in \texttt{R}.  The R-INLA package in \cite{martino2009implementing} is used along with the neural network package Keras \citep{arnold2017kerasr}. Choices of hyperparameters, the number of hidden nodes, etc., are now discussed

\begin{itemize}

\item \textbf{R-INLA} :  \\
        The R-INLA package of \cite{martino2009implementing} implements a two-dimensional stochastic partial differential equation (SPDE) approach to model spatial processes. \cite{meshchoice} discuss hyperparameter choices in this method, such as mesh-building parameters. Two crucial parameters, which determine the trianglular densities of the mesh and are generated by \texttt{inla.mesh.2d}, are \texttt{max.edge}, which controls the largest triangle edge length, and \texttt{cutoff}, which controls the minimum distance between two points. Care is needed to define a proper value of \texttt{max.edge} as this parameter depends on \texttt{cutoff}; other parameters are set to default package values. The \texttt{max.edge} parameters are set by first choosing the smaller of the range of longitude and latitude.  This is then divided by the square root of the total number of observations, then multiplied by an adapting parameter $\theta > 0$, chosen by cross validation, for each data set.  This is quantified by a parameter $\phi^*$ obeying 
        \begin{gather}
            \phi^* = \frac{ \min\left(r({\rm Longitude}), r({\rm Latitude})\right)}{\sqrt{N}} \times \theta, \nonumber
        \end{gather}
        where $r(\cdot)$ calculates the range of a vector of observations. Smaller $\theta$ induce denser triangles in the mesh; however, computations increase rapidly with smaller $\theta$ due to the increase in the number of knots.
        
    \item \textbf{Baseline Deep Neural Network} : \\
       This deep neural network trains the model using only longitude and latitude. The longitude and latitude are first scaled with min-max methods before being fed into the neural network. There are three hidden layers, all using the ReLU activation function, 100 hidden units in each layer, and one output layer with a linear activation function. A drop-out layer before each weight layer is applied with a dropout rate of $p=0.1$. Finally, a batch normalization layer is used after each hidden layer.

       The dataset is split into training and test portions for cross validation. The model is only trained on the training data; performance is evaluated on the test data. The optimizer Adam \citep{kingma2014adam} is used with a squared error loss. Early stopping is applied: a subset of training data, called a validation set, is further split to determine when to stop model training to avoid overfitting. The model is called trained when all training data has been used in the stochastic gradient descent based optimizer. At the end of each epoch, if the model fits becomes better (a smaller loss on the validation data), the model is saved and replaces the previous optimal model.    The best model is applied to the test data to evaluate model performance.
 
       In this manner, the best model among all epochs is retained. The batch size and maximum number of epochs are flexibly set for each data set after considering computational feasibility issues.
       
    \item \textbf{DeepKriging with RBFs} :  \\
            In addition to longitude and latitude, the basis functions calculated using the FRK package \cite{zammit2023introduction} are included. There may be multiple RBF resolutions; the basis functions are not scaled. Three basis function resolutions are used in our work, corresponding to the first three resolutions in the FRK package. Multiple hidden layers, each having 100 hidden nodes, are used followed by one dropout layer with a dropout rate of 0.1 and a batch normalization layer. Similarly, ReLU activation is used. The other settings are identical to  the Baseline DNN settings, but the epoch count is set to 500.  Early stopping is again employed.          
\end{itemize}

\subsection{Model Performance Scoring Rules}

Following \cite{gal2016dropout}, after training the model in each cross-validation fold, one can obtain predictive samples at each observed location. Based on these samples, the four models are judged/scored on the following criteria:

\begin{itemize}
 \item \textbf{Prediction Mean Squared Error (MSE)} : \\
    For each fold, the prediction mean squared error for the testing sample is calculated based on an out-of-sample prediction computed from 80\% of the data. Thereafter, the MSE is calculated for these predictions via
\[
{\rm MSE}(\boldsymbol{y},\boldsymbol{\hat{y}}) = 
\frac{\sum_{i=1}^N (\boldsymbol{y}-\boldsymbol{\hat{y}})^2}{N}.
\]

\item \textbf{Continuous Ranked Probability Scores (CRPSs)} : \\
The CRPS score is defined as
\[
{\rm CRPS}(F,x) = - \int_{-\infty}^\infty(F(y)-I\{ y \geq x\})^2dy.
\]
Here, $F$ is the cumulative distribution function (CDF) of the probabilistic forecast. By Lemma 2.2 of \cite{baringhaus2004new} (or identity (17) of \cite{szekely2005new}), the CRPS score has the closed form
\[
{\rm CRPS}(F,x) = \frac{1}{2}E_F[| X-X' |] - E_F[|X-x|],
\]   
where $X$ and $X^\prime$ are independent copies of a random variable with cumulative distribution function $F$ and finite first moment \citep{gneiting2007strictly}. The model with the bigger CRPS is preferred and these scores can be negative.

\item \textbf{Interval Coverage Rates (ICRs)} : \\
Level $\alpha$ credible intervals of the posterior predictive distribution at each location can be obtained via Monte Carlo methods. The interval coverage rate measures the percent of observations that fall with in the credible interval:
\[
{\rm ICR} = \frac{\sum_{i=1}^N I(x_i \in \left(L_i(\alpha), U_i(\alpha))\right)}{N}.
\]
Here, $x_i$ corresponds denotes the $i^{\rm th}$ observation, $\alpha$ the credible interval level, $L_i(\alpha)$ is the lower $\frac{\alpha}{2}$ quantile of the prediction samples, and $U_i({\alpha})$ is the upper $1-\frac{\alpha}{2}$ quantile of the prediction samples. The model with the closest ICR to $1-\alpha$ is preferred.
    
\item \textbf{Interval Scores (ISs)} : \\
The interval score is another prediction interval criterion, with smaller ISs being preferred. Two quantiles that control the upper and lower tails govern the width of the interval:
\[
U = Q\left(x, 1-\frac{\alpha}{2}\right), \quad
L = Q\left(x, \frac{\alpha}{2}\right), 
\]
where $Q$ is the quantile function. The interval score is
\[
S_\alpha^{{\rm int}}(L,U;x)= (U-L) + \frac{\alpha}{2}(L-x) I \{ x < L \} 
+\frac{\alpha}{2}(x-U) I \{ x>U \}
\]
This scoring rule has intuitive appeal and has evolved from \cite{dunsmore1968bayesian} and \cite{murphy1979probabilistic}. The predictor is rewarded for an accurate and short interval, but penalized if the observation misses the posterior interval.
\end{itemize}

\section{Simulation Study}
\label{sec:simulation}

\begin{figure}
    \centering
    \includegraphics[width=0.5\linewidth]{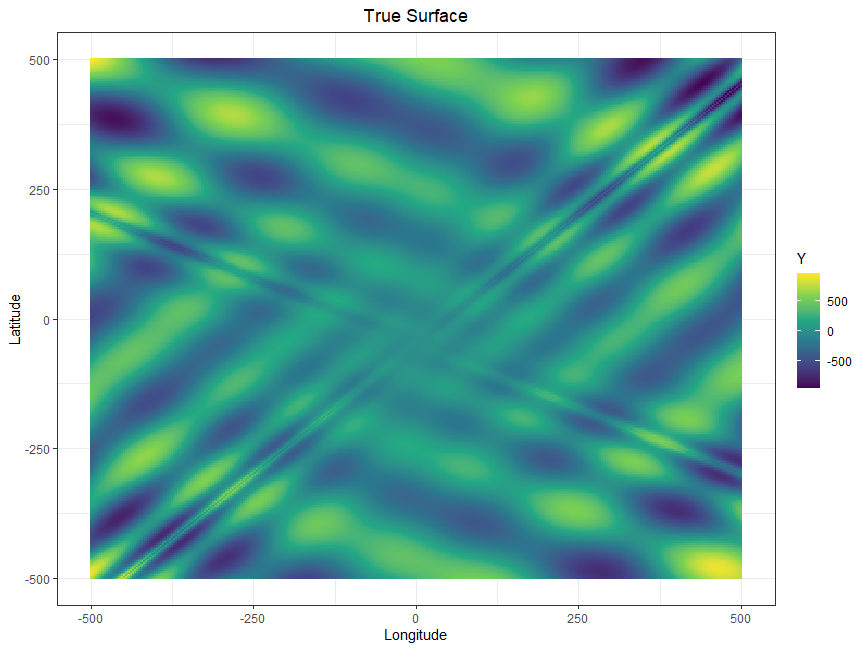}
    \caption{True Eggholder Function Values}
    \label{fig:EH}
\end{figure}

This section tests the ability of SDCNNs to predict a complex target function --- the two dimensional Eggholder function
\[
f(x_1,x_2)=-(x_2+47)\sin\left(\sqrt{\bigg|x_2+\frac{x_1}{2}+47\bigg|}\right)
-x_1\sin\left( \sqrt{\bigg|x_1-(x_2+47)\bigg|}\right),
\]
where $x_1$ and $x_2$ represent longitude and latitude. This function has multiple local maximums and minimums, as depicted in Figure \ref{fig:EH}, making prediction difficult. 

Two configurations of training data are considered.  The first uses all points in a square within a five-folded cross-validation; the second eliminates a rectangular subregion of these locations.  In the latter case, the observations within the rectangular region are predicted to evaluate procedure stability and robustness. Our $(x_1, x_2)$  points are generated by a $300 \times 300$ grid, uniformly spaced over $(x_1, x_2) \in [-500, 500] \times [-500,500]$. We scale $x_1$ and $x_2$ so that they all lie in $[0,1]^2$ on a min-max scale. Response variables are the $f(x_1,x_2)$ values (without any noise).

\subsection{Predictive Eggholder Function Performance}

\subsubsection{The Full Data Set}
 
\begin{figure}[H]
  \centering
  \includegraphics[width=0.8\textwidth]{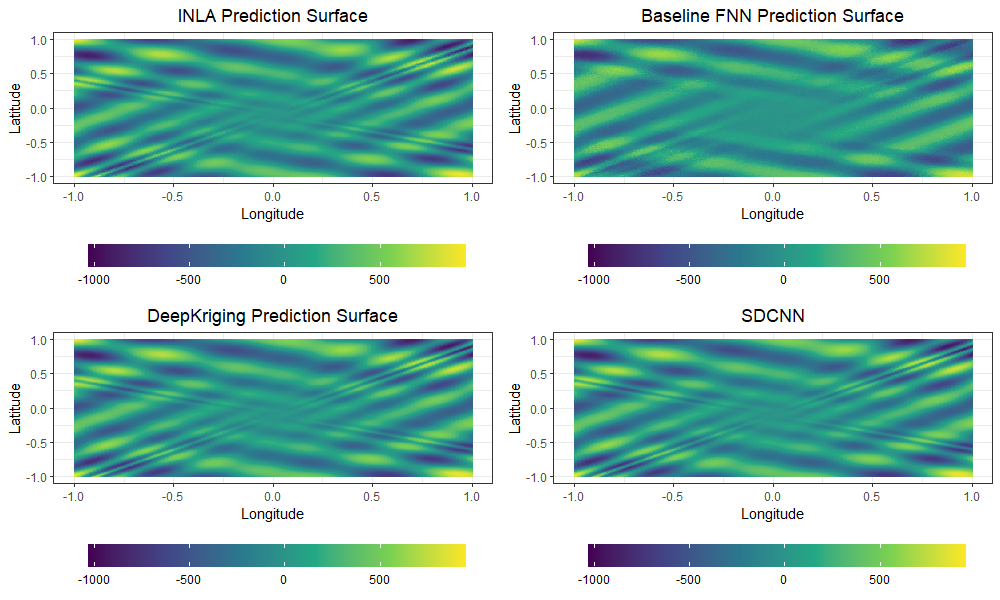}
  \caption{Predicted Mean Surface of the Eggholder Function from all Data.}
  \label{fig:pred_sur_full_eh}
\end{figure}

\begin{figure}[H]
  \centering
  \includegraphics[width=0.8\textwidth]{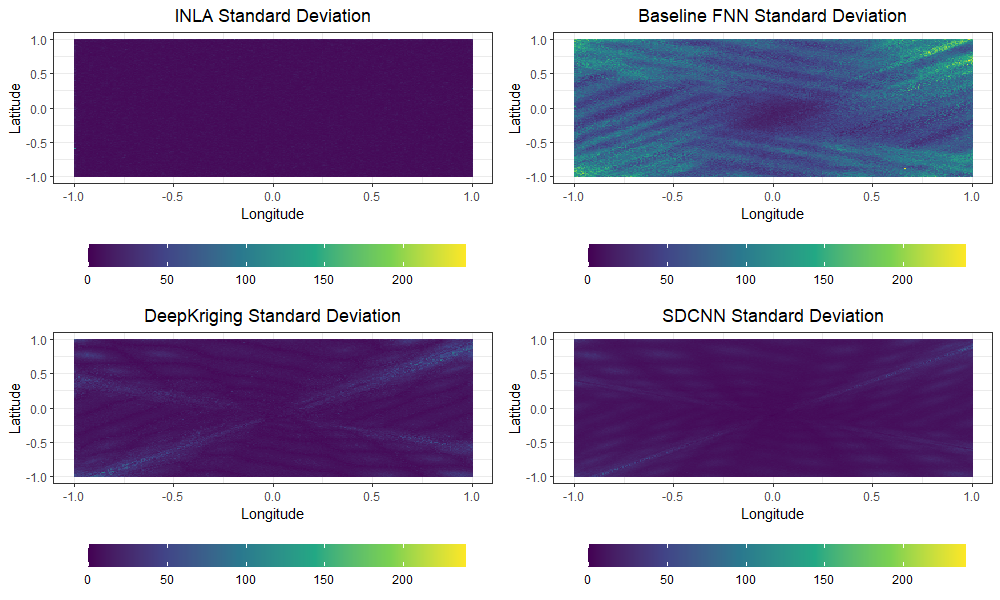}
  \caption{Prediction Standard Deviation of the Eggholder Function from all Data.}
  \label{fig:sd_sur_full_eh}
\end{figure}

First, we train a model from all observations. The predicted surface for this data is shown in Figure \ref{fig:pred_sur_full_eh}. The Baseline DNN model makes the worst predictions, ambiguously describing local Eggholder properties. Visual differences between the INLA, DeepKriging, and SDCNN methods from the plot are minor; performance scores will reveal more structure later. The uncertainty surface in Figure \ref{fig:sd_sur_full_eh} shows that the INLA method has the smallest prediction standard deviation of all models. The Baseline DNN has the largest standard deviation (with obvious biases), and the SDCNN model has a smaller standard deviation than the DeepKriging model.

Using dropout layers to assess prediction uncertainty, 100 prediction samples for the Baseline DNN, DeepKring, and SDCNN models were made for each point. For INLA, posterior predictive samples are accessibly obtained through the \texttt{R} INLA package, allowing us to calculate mean squared prediction errors, 0.95 ICRs, CRPS, and ISs for each spatial location. The scores for these four models are shown in Table \ref{tab:score_full_eh} and reveal that the SDCNN method has the smallest MSE, CRPS, and ISs. This said, the 0.95 interval coverage rate seems excessively high (not as close to 0.95 as DeepKriging).

\begin{table}[htbp]
\caption{Scores for Full Eggholder Simulation Dataset}
\label{tab:score_full_eh}
\centering
\begin{tabular}{ccccc}
  \hline
 & MSE &  CRPS & ICR(0.95) & IS(0.05) \\ 
  \hline
INLA-SPDE & 297.57 & -5.27 & 0.91 & 125.77 \\ 
  Baseline FNN & 5584.53 & -38.23 & 0.83 & 568.99 \\  
  DeepKriging & 171.64 & -5.55 & {\color{red}\textbf{0.96}} & 61.51 \\ 
  SDCNN & {\color{red}\textbf{52.68}} & {\color{red}\textbf{-3.24}} & 0.99 & {\color{red}\textbf{41.63}} \\ 
   \hline 
\hline
\end{tabular}
\end{table}

\subsubsection{The Reduced Data Set}

\begin{figure}[H]
\centering
\includegraphics[width=1\textwidth]{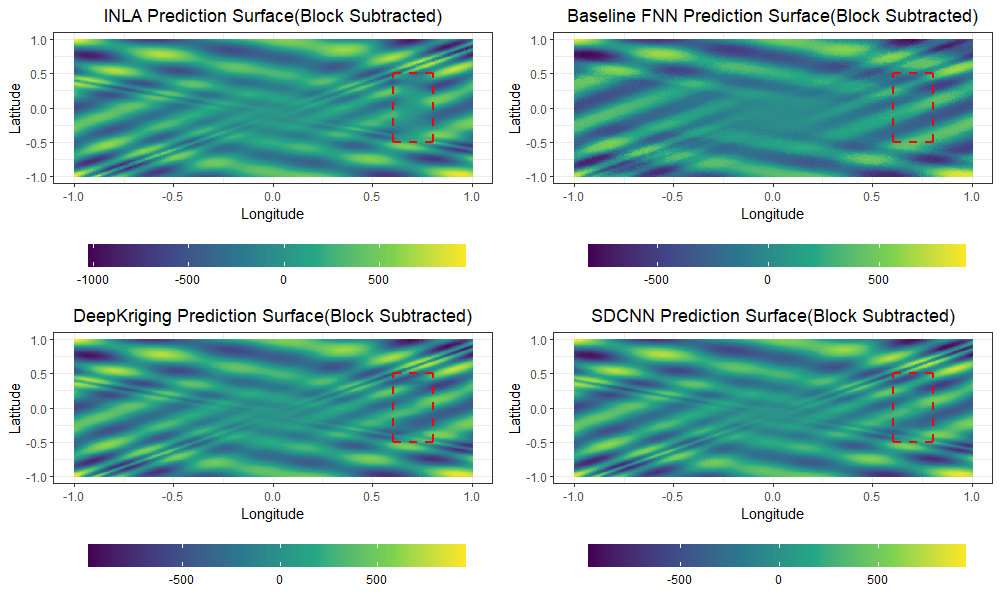}
\caption{Predicted Eggholder Surface with a Missing Rectangle.}
\label{fig:pred_sur_emp_eh}
\end{figure}

\begin{figure}[H]
  \centering
  \includegraphics[width=1\textwidth]{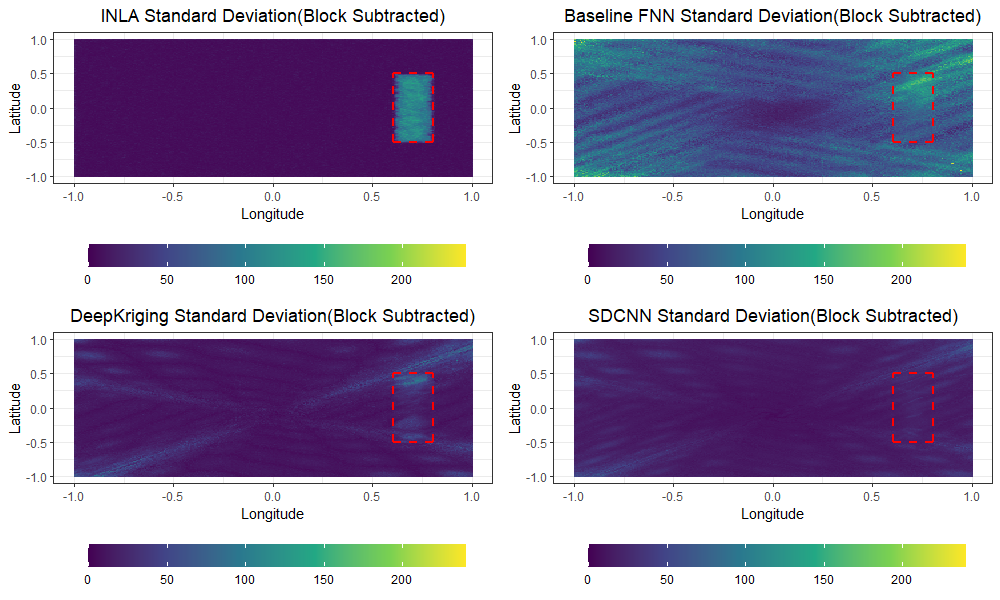}
  \caption{Standard Deviation of Eggholder Surface with a Missing Rectangle.}
  \label{fig:sd_sur_emp_eh}
\end{figure}

This case repeats the last subsection's exercise with the reduced data set. The predictive mean surface and standard deviation plots are shown in Figures \ref{fig:pred_sur_emp_eh} and \ref{fig:sd_sur_emp_eh}. For the predictive mean surface, INLA did not reproduce the ``diagonal lines" structure in the Eggholder function values.  The best predictive model is again the SDCNN, whose predictions did not overly degrade with the omitted rectangle. In contrast, the standard deviation of INLA predictions increased greatly when the rectangle was omitted. The DeepKriging method also did not significantly degrade with the omitted rectangle. Here, SDCNNs are more stable than INLA techniques when local observations are rare; moreover, SDCNNs have the best performance scores when the rectangle is not omitted.

\section{Applications}
\label{sec:application}

\subsection{Satellite Temperatures}

\begin{figure}[H]
  \centering
  \includegraphics[width=1\textwidth]{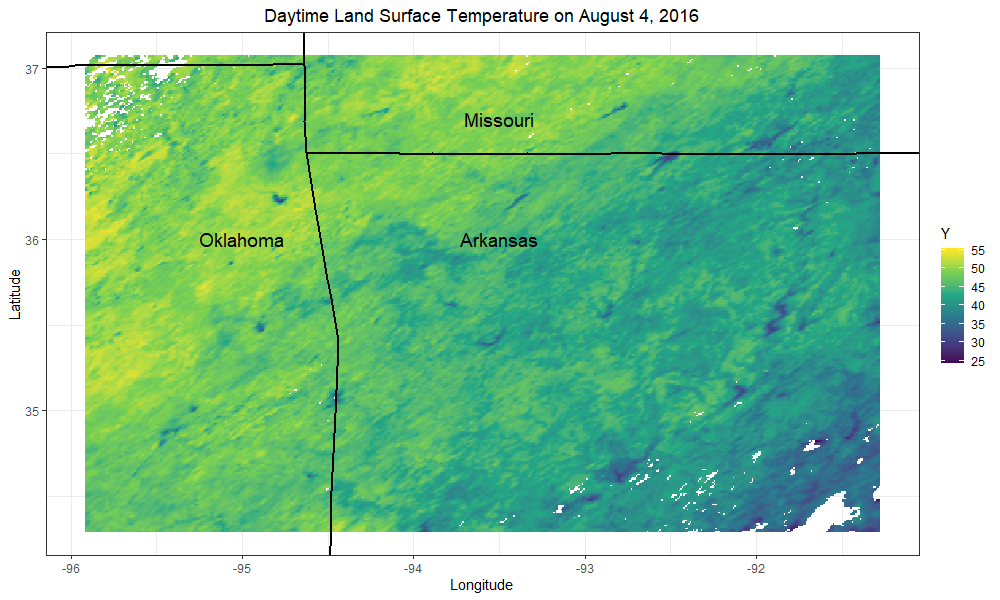}
  \caption{Observation Surface of the Satellite Data}
  \label{fig:SAT_obs}
\end{figure}

Now we apply our SDCNN method to a data set used in \cite{heaton2019case}, which contains observations on the $500 \times 300$ equally spaced grid spanning longitudes from -95.91153 degrees W to -91.28381 degrees W and latitudes from 34.29519 degrees N to 37.06811 degrees N.  About 1.1\% of the data are missing because of clouds. The $Y$ variable is the daytime land surface temperature recorded by the Terra instrument onboard the MODIS satellite on August 4, 2016 (Level-3 data). The data are available at \url{https://github.com/finnlindgren/heatoncomparison}, contain 148,309 observations, and are shown in Figure \ref{fig:SAT_obs}.

\begin{figure}[H]
  \centering
  \includegraphics[width=1\textwidth]{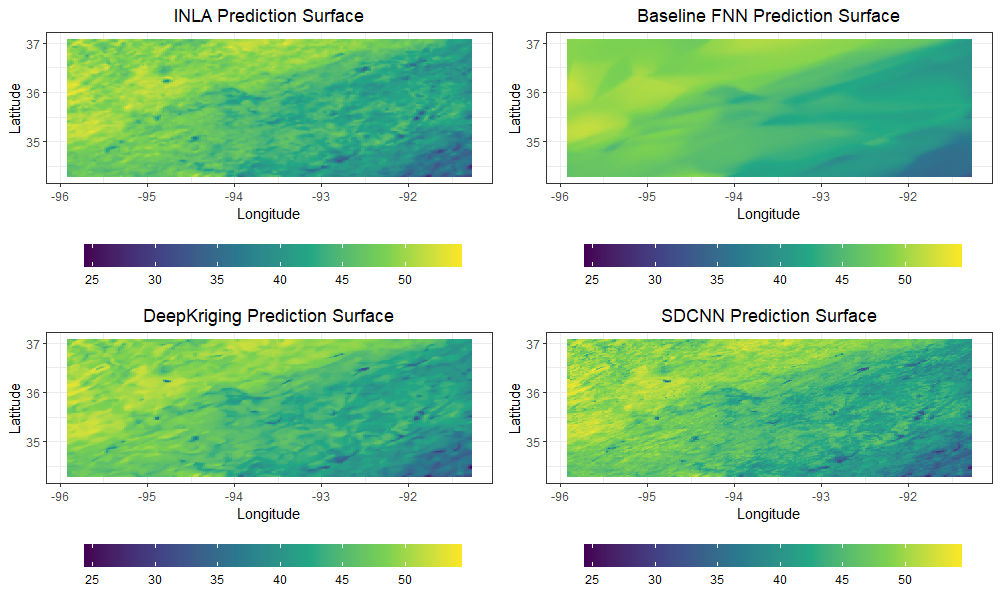}
  \caption{Predicted Surface of the Satellite Temperatures}
  \label{fig:pred_sur_sat}
\end{figure}

We generated a $200 \times 200$ grid of longitudes and latitudes over the study region and predicted the surface temperatures at all locations on the $200 \times 200$ grid.  The predicted surface of the four models above is shown in Figure \ref{fig:pred_sur_sat} over the $200 \times 200$ grid. The SDCNN method best matches the true data, efficiently capturing local variations in the surface. INLA methods also capture some of the local structure, but the image is blurry when compared to SDCNN predictive surface. The DeepKriging and Baseline DNN methods did not work well at all.

\begin{figure}[H]
  \centering
  \includegraphics[width=1\textwidth]{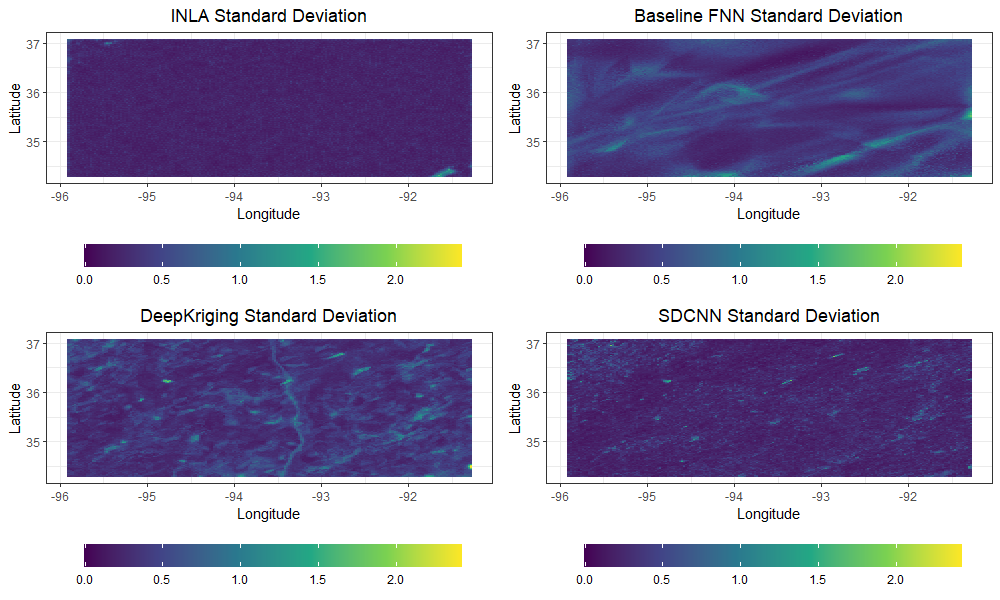}
  \caption{Uncertainty Surface of the Satellite Temperatures}
  \label{fig:SAT_sd_sur}
\end{figure}

The INLA predictions in Figure \ref{fig:SAT_sd_sur} have a large standard deviation in the bottom right corner locations where data are missing. This said, the INLA predictions have the smallest standard deviation when observations exist nearby. 

The prediction MSEs, CRPSs, ICRs, and ISs for each fold are shown in Table \ref{tab:score_sat}. The SDCNN method performs best among all models. 

\begin{table}[ht]
\caption{Scores for the Satellite Temperatures}
\label{tab:score_sat}
\centering
\begin{tabular}{ccccc}
\hline
 & MSE & CRPS & ICR(0.95) & IS(0.05) \\ 
\hline
INLA-SPDE & 1.04 & -0.65 & 0.37 & 17.64 \\ 
Baseline FNN &2.02 & -0.89 & 0.44 & 21.48 \\
DeepKriging & 0.78 & -0.54 & 0.59 & 9.99 \\
SDCNN & {\color{red}\textbf{0.31}} & {\color{red}\textbf{-0.30}} & {\color{red}\textbf{0.70}} & {\color{red}\textbf{4.77}} \\
\hline
\end{tabular}
\end{table}

In the above table, the SDCNN method is seen to perform better than the other three models, uniformly in the scoring metrics.

\begin{figure}[H]
  \centering
  \includegraphics[width=1\textwidth]{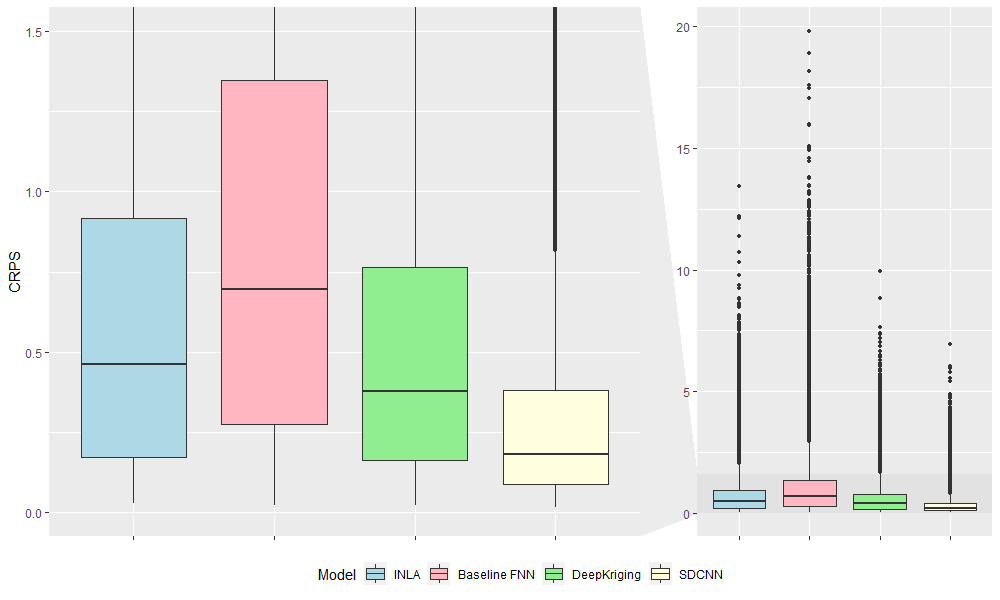}
  \caption{Boxplot of CRPS by Method for the Satellite Temperatures}
  \label{fig:crps_sat}
\end{figure}

A boxplot of the CRPSs of these four models is shown in Figure \ref{fig:crps_sat} and reveals further structure. The SDCNN method has the smallest quartile of all four models, indicating its superiority.  Perhaps more importantly, the larger outliers of the other methods are avoided in the SDCNN.

\begin{figure}[htbp]
  \centering
  \includegraphics[width=1\textwidth]{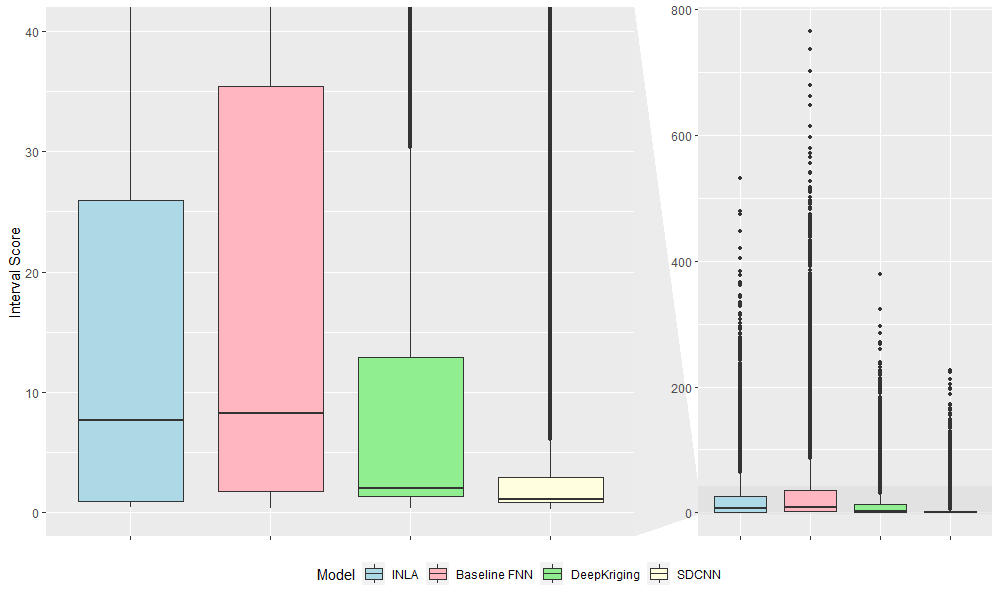}
  \caption{Boxplot of Interval Scores by Method for the Satellite Temeratures}
  \label{fig:int_sat}
\end{figure}

Figures \ref{fig:int_sat} show boxplots of ISs. The SDCNN model again performs best.

\subsection{United States Soil Carbon}

\begin{figure}[H]
  \centering
  \includegraphics[width=1\textwidth]{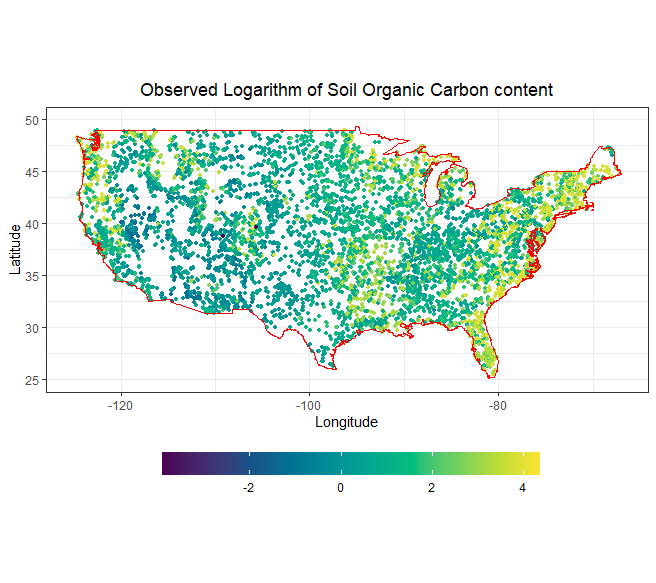}
  \caption{Observed Logarithm of the SOC Data.}
  \label{fig:soc_obs}
\end{figure}

In our previous data sets, all observations roughly lied on a lattice. Our next example involves soil carbon measurements from the conterminous United States and is more ``non-lattice". The data were extracted from the in-built R package \texttt{soilDB} created by \cite{beaudette2023package}. This package extracts soil properties from the databases of the US Department of Agriculture's Natural Resources Conservation Service (USDA-NRCS) and the National Cooperative Soil Survey (USDA-NCSS). Longitude and latitude are used as covariates; soil organic carbon (SOC) content is our response variable. The code to extract the data is on our GitHub page for this paper. The dataset contains 31,491 irregularly spaced observations as shown in Figure \ref{fig:soc_obs}. For our prediction and uncertainty surfaces, we again generate a $200 \times 200$ grid encompassing the spatial domain.

\begin{figure}[H]
\centering
\includegraphics[width=1\textwidth]{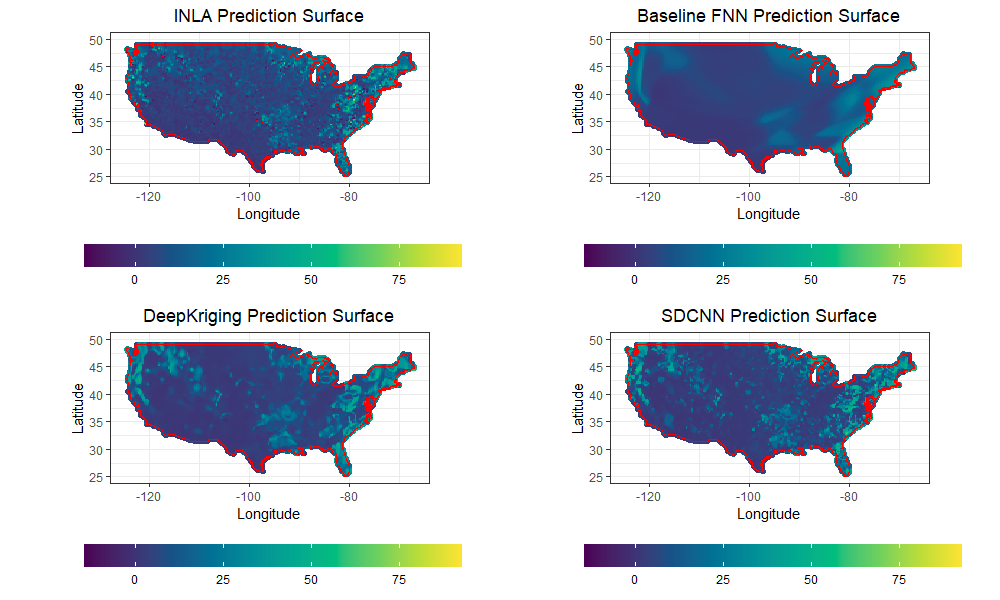}
\caption{Predictions of the SOC Data.}
\label{fig:soc_pred_sur}
\end{figure}

Our predicted surface is shown in Figure \ref{fig:soc_pred_sur}. The INLA and SDCNN surfaces more accurately predict the observations than the other two methods, presumedly because they more accurately explore local data features. Although the DeepKriging and Baseline DNN methods both indicate that the Eastern United States has larger SOCs, they perform poorly in the Central United States.

\begin{figure}[H]
  \centering
  \includegraphics[width=1\textwidth]{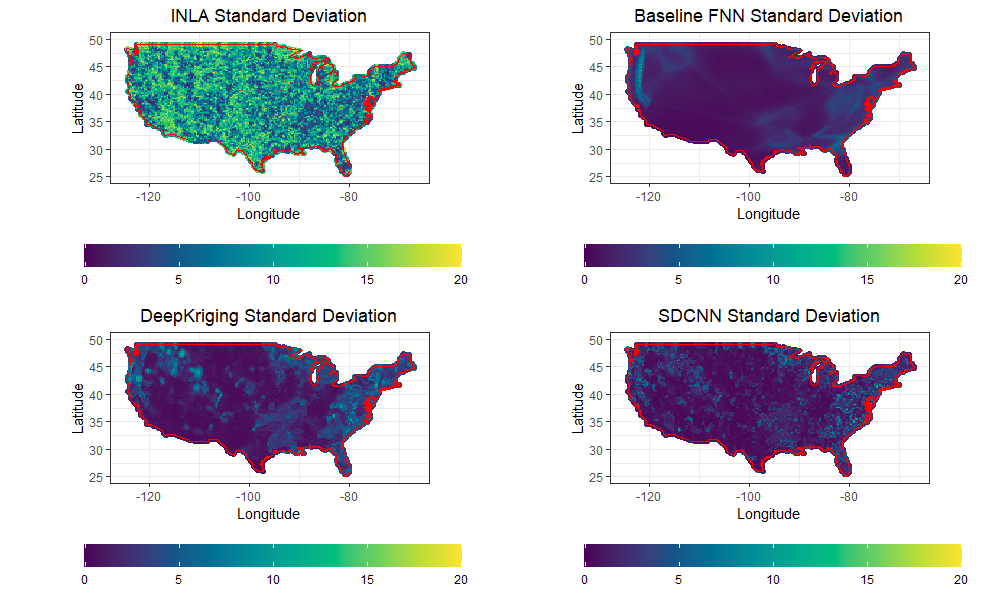}
  \caption{Uncertainty Surface for the SOC Data.}
  \label{fig:soc_sd_sur}
\end{figure}

Figure \ref{fig:soc_sd_sur} illuminates the drawbacks of INLA, which performs more poorly here than with the satellite temperatures, presumedly because the data are more non-lattice. The DeepKriging, Baseline DNN, and SDCNN methods perform similarly, but the SDCNN predictions have a more detailed local structure.

\begin{figure}[H]
  \centering
  \includegraphics[width=0.8\textwidth]{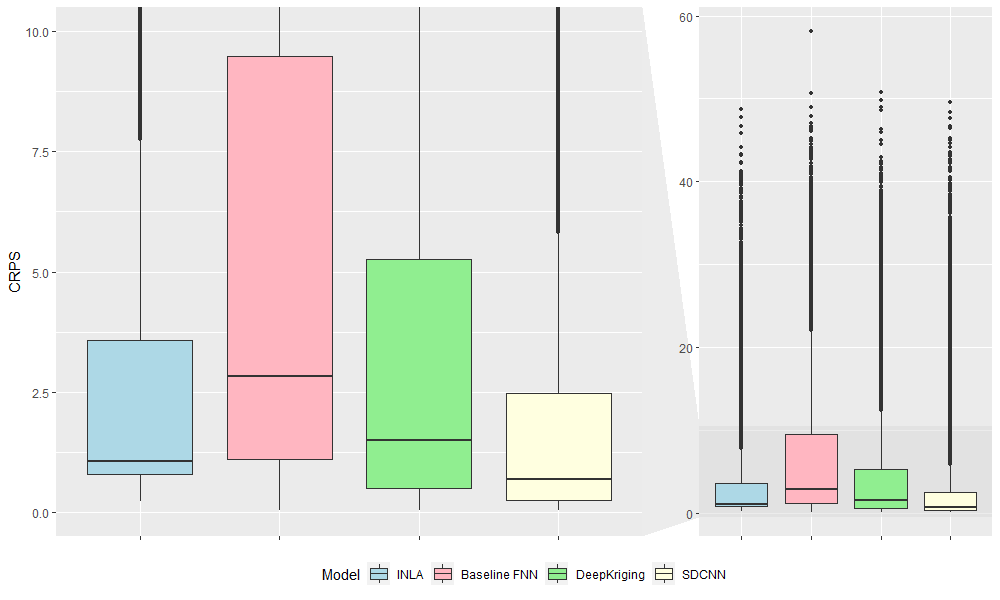}
  \caption{CRPS for the United States SOC Data.}
  \label{fig:crps_soc}
\end{figure}

The evaluation scores of these four models are shown in Table \ref{tab:soc_scores}. The SDCNN model has the best scores for all criteria except the ICRs, where INLA performs better. However, the ISs for the SDCNN better those from INLA. CRPSs for the four models are shown in Figure \ref{fig:crps_soc}. The SDCNN method again has the smallest quartiles.

\begin{table}[ht]
\caption{Evaluation Scores for the United States SOC Data.}
\label{tab:soc_scores}
\centering
\begin{tabular}{ccccc}
  \hline
 & MSE & CRPS & ICR(0.95) & IS(0.05) \\ 
  \hline
INLA-SPDE & 53.32 & -3.56 & {\color{red}\textbf{0.76}} & 82.08 \\ 
  Baseline FNN &109.08 & -5.95 & 0.32 & 166.64 \\ 
  DeepKriging & 56.09 & -3.49 & 0.56 & 79.97 \\ 
  SDCNN &  {\color{red}\textbf{43.89 }}& {\color{red}\textbf{-2.90}} & 0.60 & {\color{red}\textbf{77.81}} \\ 
   \hline
\end{tabular}
\end{table}

\section{Discussion}
\label{sec:discussion}

This research proposed a new spatial prediction model and compared it to several standard models, including INLA, Baseline DNN, and DeepKriging. We first generated RBFs by using the \texttt{FRK} package in \texttt{R} and used this information as image covariates in our model.  The SDCNN method was implemented in \texttt{Keras} in \texttt{R}, using the image covariates. Model uncertainty was addressed via dropout layers, which generate prediction samples that provide a prediction error surface and enable model evaluation scores to be calculated.

This SDCNN method was compared to other techniques in a simulation study where the response variable was the Eggholder function surface.  Two real data sets were also considered, one where locations lie primarily on a lattice and one where the observations were irregularly spaced. The SDCNN beat the other techniques in most evaluation scores, although not uniformly so. 

Some additional research avenues are evident. Future work includes developing convolutional recurrent neural networks (CRNNs) to analyze spatio-temporal models, where RBF generation is handled as here. Temporal aspects of the problem could also employ a CRNN, producing a spatio-temporal model that capably handles geographical time series data. Discrete data, particularly count structures, also merit study. This could be as straightforward as changing the activation and loss functions, or may demand a more detailed procedure involving the count's marginal distribution(s).  Tailoring the RBF generation to the problem at hand is also worthy of consideration. 

\section*{Acknowledgement}

This research was partially supported by the U.S. National Science Foundation (NSF) under NSF Grant NCSE-2215169. This article is released to inform interested parties of ongoing research and to encourage discussion. The views expressed on statistical issues are those of the authors and not those of the NSF.

\newpage

\bibliographystyle{apalike}

\begin{thebibliography}{}

\bibitem[Albawi et~al., 2017]{albawi2017understanding}
Albawi, S., Mohammed, T.~A., and Al-Zawi, S. (2017).
\newblock Understanding of a convolutional neural network.
\newblock In {\em 2017 International Conference on Engineering and Technology
  (ICET)}, pages 1--6. Ieee.

\bibitem[Araki et~al., 2015]{araki2015application}
Araki, S., Yamamoto, K., Kondo, A., et~al. (2015).
\newblock Application of regression {K}riging to air pollutant concentrations
  in {J}apan with high spatial resolution.
\newblock {\em Aerosol and Air Quality Research}, 15(1):234--241.

\bibitem[Arnold, 2017]{arnold2017kerasr}
Arnold, T.~B. (2017).
\newblock keras{R}: {R} interface to the keras deep learning library.
\newblock {\em J. Open Source Softw.}, 2(14):296.

\bibitem[Banerjee et~al., 2008]{banerjee2008gaussian}
Banerjee, S., Gelfand, A.~E., Finley, A.~O., and Sang, H. (2008).
\newblock {G}aussian predictive process models for large spatial data sets.
\newblock {\em Journal of the Royal Statistical Society Series B: Statistical
  Methodology}, 70(4):825--848.

\bibitem[Baringhaus and Franz, 2004]{baringhaus2004new}
Baringhaus, L. and Franz, C. (2004).
\newblock On a new multivariate two-sample test.
\newblock {\em Journal of Multivariate Analysis}, 88(1):190--206.

\bibitem[Beaudette et~al., 2023]{beaudette2023package}
Beaudette, D., Skovlin, J., Roecker, S., and Beaudette, M.~D. (2023).
\newblock Package ‘soildb’.

\bibitem[Bornn et~al., 2012]{bornn2012modeling}
Bornn, L., Shaddick, G., and Zidek, J.~V. (2012).
\newblock Modeling nonstationary processes through dimension expansion.
\newblock {\em Journal of the American Statistical Association},
  107(497):281--289.

\bibitem[Cellura et~al., 2008]{cellura2008wind}
Cellura, M., Cirrincione, G., Marvuglia, A., and Miraoui, A. (2008).
\newblock Wind speed spatial estimation for energy planning in {S}icily: A
  neural {K}riging application.
\newblock {\em Renewable Energy}, 33(6):1251--1266.

\bibitem[Chen et~al., 2024]{chen2020deepkriging}
Chen, W., Li, Y., Reich, B.~J., and Sun, Y. (2024).
\newblock Deep{K}riging: Spatially dependent deep neural networks for spatial
  prediction.
\newblock {\em Statistica Sinica}, 34:291--311.

\bibitem[Cracknell and Reading, 2014]{cracknell2014geological}
Cracknell, M.~J. and Reading, A.~M. (2014).
\newblock Geological mapping using remote sensing data: {A} comparison of five
  machine learning algorithms, their response to variations in the spatial
  distribution of training data and the use of explicit spatial information.
\newblock {\em Computers \& Geosciences}, 63:22--33.

\bibitem[Cressie and Johannesson, 2008]{cressie2008fixed}
Cressie, N. and Johannesson, G. (2008).
\newblock Fixed rank {K}riging for very large spatial data sets.
\newblock {\em Journal of the Royal Statistical Society Series B: Statistical
  Methodology}, 70(1):209--226.

\bibitem[Daw and Wikle, 2023]{daw2023reds}
Daw, R. and Wikle, C.~K. (2023).
\newblock Reds: Random ensemble deep spatial prediction.
\newblock {\em Environmetrics}, 34(1):e2780.

\bibitem[Dunsmore, 1968]{dunsmore1968bayesian}
Dunsmore, I. (1968).
\newblock A {B}ayesian approach to calibration.
\newblock {\em Journal of the Royal Statistical Society: Series B
  (Methodological)}, 30(2):396--405.

\bibitem[Finley et~al., 2020]{finley2020spnngp}
Finley, A.~O., Datta, A., and Banerjee, S. (2020).
\newblock sp{N}{N}{G}{P} {R} package for nearest neighbor {G}aussian process
  models.
\newblock {\em arXiv preprint arXiv:2001.09111}.

\bibitem[Gal and Ghahramani, 2016]{gal2016dropout}
Gal, Y. and Ghahramani, Z. (2016).
\newblock Dropout as a {B}ayesian approximation: {R}epresenting model
  uncertainty in deep learning.
\newblock In {\em International Conference on Machine Learning}, pages
  1050--1059. PMLR.

\bibitem[Gneiting and Raftery, 2007]{gneiting2007strictly}
Gneiting, T. and Raftery, A.~E. (2007).
\newblock Strictly proper scoring rules, prediction, and estimation.
\newblock {\em Journal of the American statistical Association},
  102(477):359--378.

\bibitem[Gramacy and Lee, 2008]{gramacy2008bayesian}
Gramacy, R.~B. and Lee, H. K.~H. (2008).
\newblock {B}ayesian treed {G}aussian process models with an application to
  computer modeling.
\newblock {\em Journal of the American Statistical Association},
  103(483):1119--1130.

\bibitem[Heaton et~al., 2019]{heaton2019case}
Heaton, M.~J., Datta, A., Finley, A.~O., Furrer, R., Guinness, J., Guhaniyogi,
  R., Gerber, F., Gramacy, R.~B., Hammerling, D., Katzfuss, M., et~al. (2019).
\newblock A case study competition among methods for analyzing large spatial
  data.
\newblock {\em Journal of Agricultural, Biological and Environmental
  Statistics}, 24:398--425.

\bibitem[Higdon, 1998]{higdon1998process}
Higdon, D. (1998).
\newblock A process-convolution approach to modelling temperatures in the
  {N}orth {A}tlantic {O}cean.
\newblock {\em Environmental and Ecological Statistics}, 5:173--190.

\bibitem[Hooshmand et~al., 2011]{hooshmand2011application}
Hooshmand, A., Delghandi, M., Izadi, A., Aali, K.~A., et~al. (2011).
\newblock Application of {K}riging and co{K}riging in spatial estimation of
  groundwater quality parameters.
\newblock {\em African Journal of Agricultural Research}, 6(14):3402--3408.

\bibitem[Kim et~al., 2005]{kim2005analyzing}
Kim, H.-M., Mallick, B.~K., and Holmes, C.~C. (2005).
\newblock Analyzing nonstationary spatial data using piecewise {G}aussian
  processes.
\newblock {\em Journal of the American Statistical Association},
  100(470):653--668.

\bibitem[Kingma and Ba, 2014]{kingma2014adam}
Kingma, D.~P. and Ba, J. (2014).
\newblock Adam: A method for stochastic optimization.
\newblock {\em arXiv preprint arXiv:1412.6980}.

\bibitem[Kirsner and Sans{\'o}, 2020]{kirsner2020multi}
Kirsner, D. and Sans{\'o}, B. (2020).
\newblock Multi-scale shotgun stochastic search for large spatial datasets.
\newblock {\em Computational Statistics \& Data Analysis}, 146:106931.

\bibitem[Krizhevsky et~al., 2012]{krizhevsky2012imagenet}
Krizhevsky, A., Sutskever, I., and Hinton, G.~E. (2012).
\newblock Imagenet classification with deep convolutional neural networks.
\newblock {\em Advances in Neural Information Processing Systems}, 25.

\bibitem[LeCun et~al., 1995]{lecun1995convolutional}
LeCun, Y., Bengio, Y., et~al. (1995).
\newblock Convolutional networks for images, speech, and time series.
\newblock {\em The handbook of brain theory and neural networks},
  3361(10):1995.

\bibitem[Lemos and Sans{\'o}, 2009]{lemos2009spatio}
Lemos, R.~T. and Sans{\'o}, B. (2009).
\newblock A spatio-temporal model for mean, anomaly, and trend fields of
  {N}orth {A}tlantic sea surface temperature.
\newblock {\em Journal of the American Statistical Association},
  104(485):5--18.

\bibitem[Lemos and Sans{\'o}, 2012]{lemos2012conditionally}
Lemos, R.~T. and Sans{\'o}, B. (2012).
\newblock Conditionally linear models for non-homogeneous spatial random
  fields.
\newblock {\em Statistical Methodology}, 9(1-2):275--284.

\bibitem[Lindgren et~al., 2011]{lindgren2011explicit}
Lindgren, F., Rue, H., and Lindstr{\"o}m, J. (2011).
\newblock An explicit link between {G}aussian fields and {G}aussian {M}arkov
  random fields: the stochastic partial differential equation approach.
\newblock {\em Journal of the Royal Statistical Society Series B: Statistical
  Methodology}, 73(4):423--498.

\bibitem[Martino and Rue, 2009]{martino2009implementing}
Martino, S. and Rue, H. (2009).
\newblock Implementing approximate {B}ayesian inference using {I}ntegrated
  {N}ested {L}aplace {A}pproximation: A manual for the inla program.
\newblock {\em Department of Mathematical Sciences, NTNU, Norway}.

\bibitem[Murphy and Winkler, 1979]{murphy1979probabilistic}
Murphy, A.~H. and Winkler, R.~L. (1979).
\newblock Probabilistic temperature forecasts: {T}he case for an operational
  program.
\newblock {\em Bulletin of the American Meteorological Society}, 60(1):12--19.

\bibitem[Nag et~al., 2023a]{nag2023bivariate}
Nag, P., Sun, Y., and Reich, B.~J. (2023a).
\newblock Bivariate {D}eep{K}riging for large-scale spatial interpolation of
  wind fields.
\newblock {\em arXiv preprint arXiv:2307.08038}.

\bibitem[Nag et~al., 2023b]{nag2023spatio}
Nag, P., Sun, Y., and Reich, B.~J. (2023b).
\newblock Spatio-temporal {D}eep{K}riging for interpolation and probabilistic
  forecasting.
\newblock {\em Spatial Statistics}, 57:100773.

\bibitem[Neal, 2012]{neal2012bayesian}
Neal, R.~M. (2012).
\newblock {\em Bayesian learning for neural networks}, volume 118.
\newblock Springer Science \& Business Media.

\bibitem[Nychka et~al., 2015]{nychka2015multiresolution}
Nychka, D., Bandyopadhyay, S., Hammerling, D., Lindgren, F., and Sain, S.
  (2015).
\newblock A multiresolution {G}aussian process model for the analysis of large
  spatial datasets.
\newblock {\em Journal of Computational and Graphical Statistics},
  24(2):579--599.

\bibitem[O'Shea and Nash, 2015]{o2015introduction}
O'Shea, K. and Nash, R. (2015).
\newblock An introduction to convolutional neural networks.
\newblock {\em arXiv preprint arXiv:1511.08458}.

\bibitem[Righetto et~al., 2020]{meshchoice}
Righetto, A.~J., Faes, C., Vandendijck, Y., and Jr., P. J.~R. (2020).
\newblock On the choice of the mesh for the analysis of geostatistical data
  using r-inla.
\newblock {\em Communications in Statistics - Theory and Methods},
  49(1):203--220.

\bibitem[Rue et~al., 2009]{rue2009approximate}
Rue, H., Martino, S., and Chopin, N. (2009).
\newblock Approximate {B}ayesian inference for latent {G}aussian models by
  using integrated nested laplace approximations.
\newblock {\em Journal of the Royal Statistical Society Series B: Statistical
  Methodology}, 71(2):319--392.

\bibitem[Rumelhart et~al., 1986]{rumelhart1986learning}
Rumelhart, D.~E., Hinton, G.~E., and Williams, R.~J. (1986).
\newblock Learning representations by back-propagating errors.
\newblock {\em nature}, 323(6088):533--536.

\bibitem[Sampson and Guttorp, 1992]{sampson1992nonparametric}
Sampson, P.~D. and Guttorp, P. (1992).
\newblock Nonparametric estimation of nonstationary spatial covariance
  structure.
\newblock {\em Journal of the American Statistical Association},
  87(417):108--119.

\bibitem[Schmidt and O'Hagan, 2003]{schmidt2003bayesian}
Schmidt, A.~M. and O'Hagan, A. (2003).
\newblock {B}ayesian inference for non-stationary spatial covariance structure
  via spatial deformations.
\newblock {\em Journal of the Royal Statistical Society Series B: Statistical
  Methodology}, 65(3):743--758.

\bibitem[Sharma et~al., 2017]{sharma2017activation}
Sharma, S., Sharma, S., and Athaiya, A. (2017).
\newblock Activation functions in neural networks.
\newblock {\em Towards Data Sci}, 6(12):310--316.

\bibitem[Srivastava et~al., 2014]{srivastava2014dropout}
Srivastava, N., Hinton, G., Krizhevsky, A., Sutskever, I., and Salakhutdinov,
  R. (2014).
\newblock Dropout: a simple way to prevent neural networks from overfitting.
\newblock {\em The Journal of Machine Learning Research}, 15(1):1929--1958.

\bibitem[Sz{\'e}kely and Rizzo, 2005]{szekely2005new}
Sz{\'e}kely, G.~J. and Rizzo, M.~L. (2005).
\newblock A new test for multivariate normality.
\newblock {\em Journal of Multivariate Analysis}, 93(1):58--80.

\bibitem[Wikle and Zammit-Mangion, 2023]{wikle2023statistical}
Wikle, C.~K. and Zammit-Mangion, A. (2023).
\newblock Statistical deep learning for spatial and spatiotemporal data.
\newblock {\em Annual Review of Statistics and Its Application},
  10(1):247--270.

\bibitem[Zammit-Mangion and Cressie, 2023]{zammit2023introduction}
Zammit-Mangion, A. and Cressie, N. (2023).
\newblock Introduction to {F}ixed {R}ank {K}riging: The r package.

\bibitem[Zammit-Mangion et~al., 2024]{zammit2023spatial}
Zammit-Mangion, A., Kaminski, M.~D., Tran, B.-H., Filippone, M., and Cressie,
  N. (2024).
\newblock Spatial bayesian neural networks.
\newblock {\em Spatial Statistics}, page 100825.

\bibitem[Zammit-Mangion et~al., 2022]{zammit2022deep}
Zammit-Mangion, A., Ng, T. L.~J., Vu, Q., and Filippone, M. (2022).
\newblock Deep compositional spatial models.
\newblock {\em Journal of the American Statistical Association},
  117(540):1787--1808.

\bibitem[Zhan and Datta, 2024]{zhan2023neural}
Zhan, W. and Datta, A. (2024).
\newblock Neural networks for geospatial data.
\newblock {\em Journal of the American Statistical Association}, 0(0):1--21.

\bibitem[Zhang et~al., 2023]{zhang2023efficient}
Zhang, J., Ju, Y., Mu, B., Zhong, R., and Chen, T. (2023).
\newblock An efficient implementation for spatial--temporal {G}aussian process
  regression and its applications.
\newblock {\em Automatica}, 147:110679.

\end{thebibliography}

\newpage

\appendix

\section{Model Visualization}

This Appendix shows how our neural networks in this paper were structured, including the Baseline DNN, DeepKriging, and SDCNN models.

\subsection{Baseline DNN}
\begin{figure}[htbp]
\begin{center}
\begin{tikzpicture}[scale = 0.8]

    \draw[rounded corners,  draw=black] (1, 5) rectangle (5, 4);
    \node at (3, 3.5+1) {\textbf{Coords}};
    \draw[->, >=stealth,line width=1.5pt] (3, 3+1) -- (3,2+1);
    
    \draw[rounded corners,  draw=black] (-1, 0) rectangle (7, 2+1);
    \draw[draw=black] (-1, 0.75) -- (7, 0.75);
    \draw[draw=black] (-1, 1.5) -- (7, 1.5);
    \draw[draw=black] (-1, 2.25) -- (7, 2.25);
    \fill[rounded corners, darkred] (0-1, 0.75) rectangle (6+1, 0);
    \fill[rounded corners, darkblue] (0-1, 2+1) rectangle (6+1, 2.25);
    \node[text=white] at (3, 0.375) {\textbf{Activated by: ReLU}};
    \node at (3, 0.375+0.75+0.75) {\textbf{Kernel: 2 $\times$100}};
    \node at (3, 0.375+0.75) {\textbf{Bias: 100}};
    \node[text=white] at (3, 3-0.375) {\textbf{Input Layer(Fully Connected)}};
    \draw[->, >=stealth,line width=1.5pt] (3, 0) -- (3,-2);
    \draw[->, >=stealth,line width=1.5pt] (7.5, -1) -- (3,-1);
    \draw[->, >=stealth,line width=1.5pt] (1.5, -1) -- (3,-1);
    \draw[rounded corners, draw=black](7.5,0) rectangle (10.5,-2);
    \fill[rounded corners, yellow!60](7.5,0) rectangle (10.5,-1);
    \node at (9, -0.5) {\textbf{Dropout}};
    \node at (9, -1.5) {\textbf{Rate = 0.1}};
    \draw[rounded corners, draw=black](-4.5,-0.5) rectangle (1.5,-1.5);
    \node at (-1.5, -1) {\textbf{Batch Normalization}};

    \draw[rounded corners,  draw=black] (0, -5) rectangle (6, -2);
    \draw[draw=black] (0, -2-0.75) -- (6, -2-0.75);
    \draw[draw=black] (0, -2-1.5) -- (6, -2-1.5);
    \draw[draw=black] (0, -2-2.25) -- (6, -2-2.25);
    \fill[rounded corners, darkred] (0, -2-2.25) rectangle (6, -5);
    \fill[rounded corners, darkblue] (0, -2) rectangle (6, -2-0.75);
    
    \node[text=white] at (3, 0.375-5) {\textbf{Activated by: ReLU}};
    \node at (3, 0.375+0.75+0.75-5) {\textbf{Kernel: 100$\times$100}};
    \node at (3, 0.375+0.75-5) {\textbf{Bias: 100}};
    \node[text=white] at (3, 3-0.375-5) {\textbf{Fully Connected Layer}};
    
    \draw[->, >=stealth,line width=1.5pt] (3, 0-5) -- (3,-2-5);
    \draw[->, >=stealth,line width=1.5pt] (7.5, -1-5) -- (3,-1-5);
    \draw[->, >=stealth,line width=1.5pt] (1.5, -1-5) -- (3,-1-5);
    \draw[rounded corners, draw=black](7.5,0-5) rectangle (10.5,-2-5);
    \fill[rounded corners, yellow!60](7.5,0-5) rectangle (10.5,-1-5);
    \node at (9, -0.5-5) {\textbf{Dropout}};
    \node at (9, -1.5-5) {\textbf{Rate = 0.1}};
    \draw[rounded corners, draw=black](-4.5,-0.5-5) rectangle (1.5,-1.5-5);
    \node at (-1.5, -1-5) {\textbf{Batch Normalization}};

    \draw[rounded corners,  draw=black] (0, -5-5) rectangle (6, -2-5);
    \draw[draw=black] (0, -2-0.75-5) -- (6, -2-0.75-5);
    \draw[draw=black] (0, -2-1.5-5) -- (6, -2-1.5-5);
    \draw[draw=black] (0, -2-2.25-5) -- (6, -2-2.25-5);
    \fill[rounded corners, darkred] (0, -2-2.25-5) rectangle (6, -5-5);
    \fill[rounded corners, darkblue] (0, -2-5) rectangle (6, -2-0.75-5);
    
    \node[text=white] at (3, 0.375-5-5) {\textbf{Activated by: ReLU}};
    \node at (3, 0.375+0.75+0.75-5-5) {\textbf{Kernel: 100$\times$100}};
    \node at (3, 0.375+0.75-5-5) {\textbf{Bias: 100}};
    \node[text=white] at (3, 3-0.375-5-5) {\textbf{Fully Connected Layer}};
    
    \draw[->, >=stealth,line width=1.5pt] (3, 0-5-5) -- (3,-2-5-5);
    \draw[->, >=stealth,line width=1.5pt] (7.5, -1-5-5) -- (3,-1-5-5);
    \draw[->, >=stealth,line width=1.5pt] (1.5, -1-5-5) -- (3,-1-5-5);
    \draw[rounded corners, draw=black](7.5,0-5-5) rectangle (10.5,-2-5-5);
    \fill[rounded corners, yellow!60](7.5,0-5-5) rectangle (10.5,-1-5-5);
    \node at (9, -0.5-5-5) {\textbf{Dropout}};
    \node at (9, -1.5-5-5) {\textbf{Rate = 0.1}};
    \draw[rounded corners, draw=black](-4.5,-0.5-5-5) rectangle (1.5,-1.5-5-5);
    \node at (-1.5, -1-5-5) {\textbf{Batch Normalization}};

    \draw[rounded corners,  draw=black] (0, -5-5-5) rectangle (6, -2-5-5);
    \draw[draw=black] (0, -2-0.75-5-5) -- (6, -2-0.75-5-5);
    \draw[draw=black] (0, -2-1.5-5-5) -- (6, -2-1.5-5-5);
    \draw[draw=black] (0, -2-2.25-5-5) -- (6, -2-2.25-5-5);
    \fill[rounded corners, darkred] (0, -2-2.25-5-5) rectangle (6, -5-5-5);
    \fill[rounded corners, darkblue] (0, -2-5-5) rectangle (6, -2-0.75-5-5);
    
    \node[text=white] at (3, 0.375-5-5-5) {\textbf{Activated by: ReLU}};
    \node at (3, 0.375+0.75+0.75-5-5-5) {\textbf{Kernel: 100$\times$100}};
    \node at (3, 0.375+0.75-5-5-5) {\textbf{Bias: 100}};
    \node[text=white] at (3, 3-0.375-5-5-5) {\textbf{Fully Connected Layer}};
     \draw[->, >=stealth,line width=1.5pt] (3, 0-5-5-5) -- (3,-1-5-5-5);

    \draw[rounded corners,  draw=black] (0, -5-5-5-4) rectangle (6, -2-5-5-4);
    \draw[draw=black] (0, -2-0.75-5-5-4) -- (6, -2-0.75-5-5-4);
    \draw[draw=black] (0, -2-1.5-5-5-4) -- (6, -2-1.5-5-5-4);
    \draw[draw=black] (0, -2-2.25-5-5-4) -- (6, -2-2.25-5-5-4);
    \fill[rounded corners, darkred] (0, -2-2.25-5-5-4) rectangle (6, -5-5-5-4);
    \fill[rounded corners, darkblue] (0, -2-5-5-4) rectangle (6, -2-0.75-5-5-4);
    
    \node[text=white] at (3, 0.375-5-5-5-4) {\textbf{Activated by: Idendity}};
    \node at (3, 0.375+0.75+0.75-5-5-5-4) {\textbf{Kernel: 100$\times$1}};
    \node at (3, 0.375+0.75-5-5-5-4) {\textbf{Bias: 1}};
    \node[text=white] at (3, 3-0.375-5-5-5-4) {\textbf{Output Layer}};
\end{tikzpicture}
\end{center}
 \caption{Baseline DNN Model Visualization.}
 \label{model:dnn}
\end{figure}
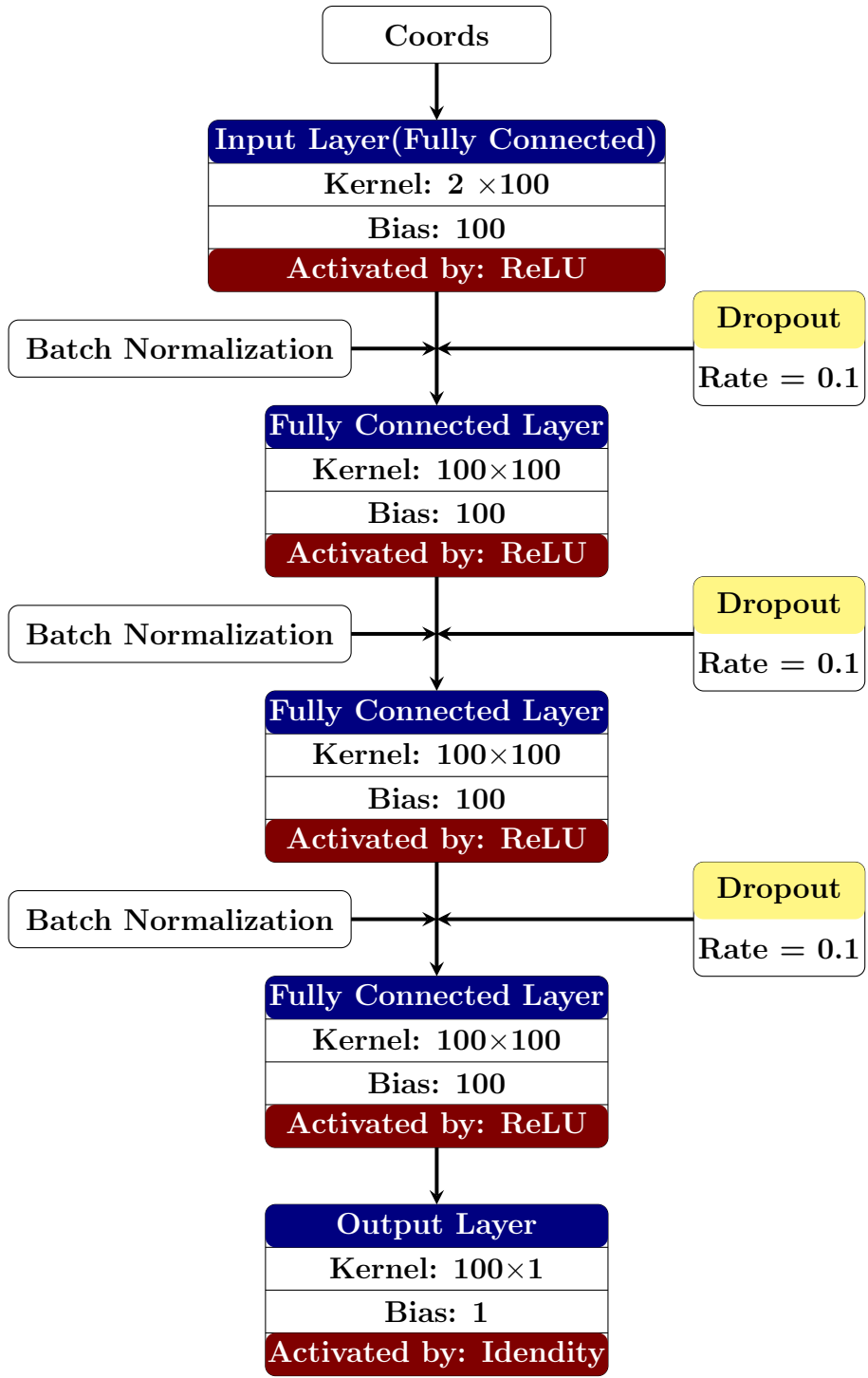

This model only uses longitude and latitude as covariates.  The model is depicted in the following graphic \ref{model:dnn}.

\subsection{DeepKriging}
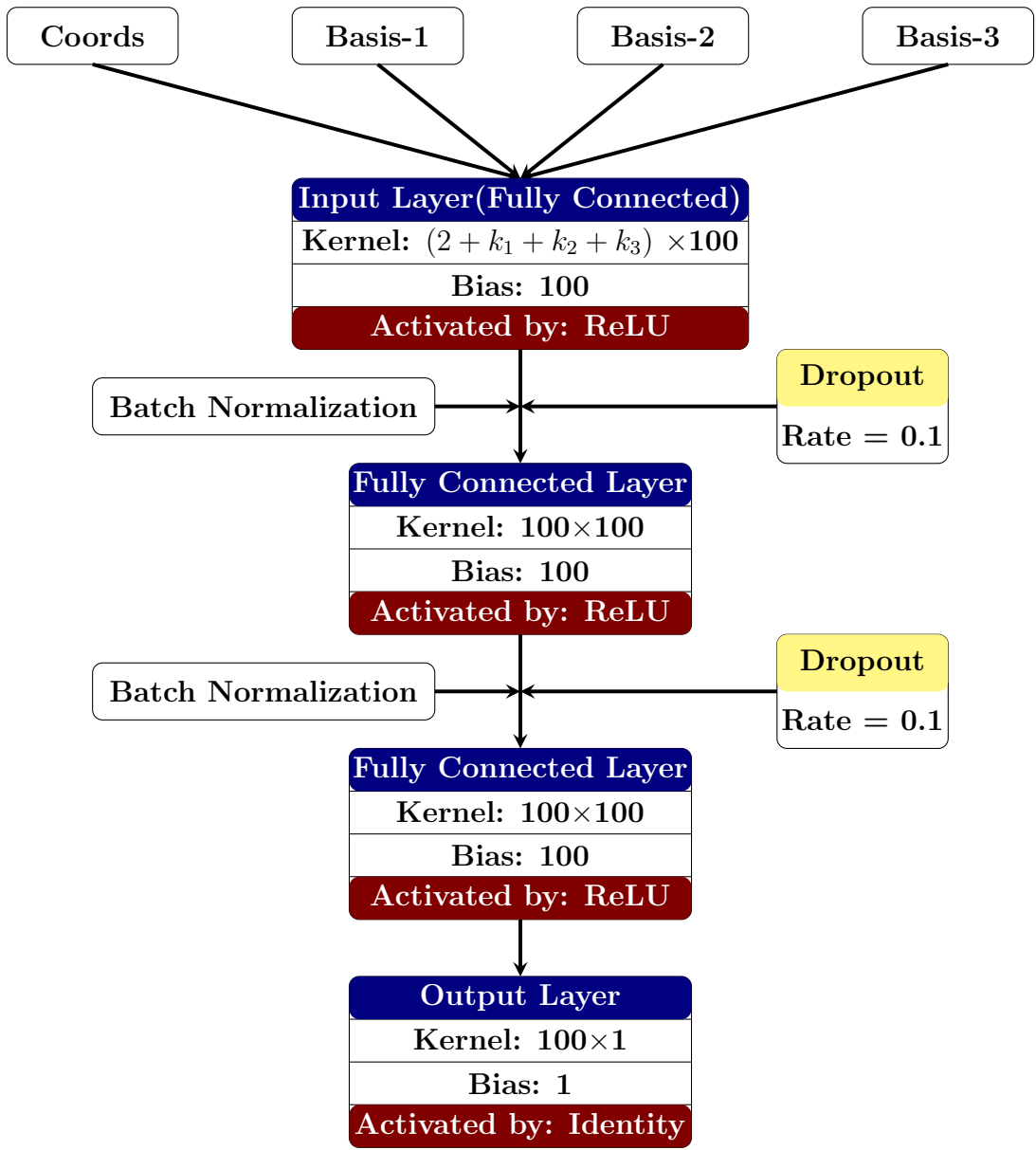
\begin{figure}
    \begin{center}
\begin{tikzpicture}[scale = 0.8]
    \draw[rounded corners,  draw=black] (-6, 5+1) rectangle (-3, 4+1);
    \draw[rounded corners,  draw=black] (12, 5+1) rectangle (9, 4+1);
    \draw[rounded corners,  draw=black] (-1, 5+1) rectangle (2, 4+1);
    \draw[rounded corners,  draw=black] (4, 5+1) rectangle (7, 4+1);
    \node at (-4.5, 4.5+1) {\textbf{Coords}};
    \node at (-4.5+5, 4.5+1) {\textbf{Basis-1}};
    \node at (-4.5+5+5, 4.5+1) {\textbf{Basis-2}};
    \node at (-4.5+5+5+5, 4.5+1) {\textbf{Basis-3}};
    
    \draw[->, >=stealth,line width=1.5pt] (-4.5,4+1) -- (3,2+1);
    \draw[->, >=stealth,line width=1.5pt] (-4.5+5,4+1) -- (3,2+1);
    \draw[->, >=stealth,line width=1.5pt] (-4.5+5+5,4+1) -- (3,2+1);
    \draw[->, >=stealth,line width=1.5pt] (-4.5+5+5+5,4+1) -- (3,2+1);

    \draw[rounded corners,  draw=black] (-1, 0) rectangle (7, 2+1);
    \draw[draw=black] (-1, 0.75) -- (7, 0.75);
    \draw[draw=black] (-1, 1.5) -- (7, 1.5);
    \draw[draw=black] (-1, 2.25) -- (7, 2.25);
    \fill[rounded corners, darkred] (0-1, 0.75) rectangle (6+1, 0);
    \fill[rounded corners, darkblue] (0-1, 2+1) rectangle (6+1, 2.25);
    \node[text=white] at (3, 0.375) {\textbf{Activated by: ReLU}};
    \node at (3, 0.375+0.75+0.75) {\textbf{Kernel: $(2+k_1+k_2+k_3)$ $\times$100}};
    \node at (3, 0.375+0.75) {\textbf{Bias: 100}};
    \node[text=white] at (3, 3-0.375) {\textbf{Input Layer(Fully Connected)}};
    \draw[->, >=stealth,line width=1.5pt] (3, 0) -- (3,-2);
    \draw[->, >=stealth,line width=1.5pt] (7.5, -1) -- (3,-1);
    \draw[->, >=stealth,line width=1.5pt] (1.5, -1) -- (3,-1);
    \draw[rounded corners, draw=black](7.5,0) rectangle (10.5,-2);
    \fill[rounded corners, yellow!60](7.5,0) rectangle (10.5,-1);
    \node at (9, -0.5) {\textbf{Dropout}};
    \node at (9, -1.5) {\textbf{Rate = 0.1}};
    \draw[rounded corners, draw=black](-4.5,-0.5) rectangle (1.5,-1.5);
    \node at (-1.5, -1) {\textbf{Batch Normalization}};

    \draw[rounded corners,  draw=black] (0, -5) rectangle (6, -2);
    \draw[draw=black] (0, -2-0.75) -- (6, -2-0.75);
    \draw[draw=black] (0, -2-1.5) -- (6, -2-1.5);
    \draw[draw=black] (0, -2-2.25) -- (6, -2-2.25);
    \fill[rounded corners, darkred] (0, -2-2.25) rectangle (6, -5);
    \fill[rounded corners, darkblue] (0, -2) rectangle (6, -2-0.75);
    
    \node[text=white] at (3, 0.375-5) {\textbf{Activated by: ReLU}};
    \node at (3, 0.375+0.75+0.75-5) {\textbf{Kernel: 100$\times$100}};
    \node at (3, 0.375+0.75-5) {\textbf{Bias: 100}};
    \node[text=white] at (3, 3-0.375-5) {\textbf{Fully Connected Layer}};
    
    \draw[->, >=stealth,line width=1.5pt] (3, 0-5) -- (3,-2-5);
    \draw[->, >=stealth,line width=1.5pt] (7.5, -1-5) -- (3,-1-5);
    \draw[->, >=stealth,line width=1.5pt] (1.5, -1-5) -- (3,-1-5);
    \draw[rounded corners, draw=black](7.5,0-5) rectangle (10.5,-2-5);
    \fill[rounded corners, yellow!60](7.5,0-5) rectangle (10.5,-1-5);
    \node at (9, -0.5-5) {\textbf{Dropout}};
    \node at (9, -1.5-5) {\textbf{Rate = 0.1}};
    \draw[rounded corners, draw=black](-4.5,-0.5-5) rectangle (1.5,-1.5-5);
    \node at (-1.5, -1-5) {\textbf{Batch Normalization}};

    \draw[rounded corners,  draw=black] (0, -5-5) rectangle (6, -2-5);
    \draw[draw=black] (0, -2-0.75-5) -- (6, -2-0.75-5);
    \draw[draw=black] (0, -2-1.5-5) -- (6, -2-1.5-5);
    \draw[draw=black] (0, -2-2.25-5) -- (6, -2-2.25-5);
    \fill[rounded corners, darkred] (0, -2-2.25-5) rectangle (6, -5-5);
    \fill[rounded corners, darkblue] (0, -2-5) rectangle (6, -2-0.75-5);
    
    \node[text=white] at (3, 0.375-5-5) {\textbf{Activated by: ReLU}};
    \node at (3, 0.375+0.75+0.75-5-5) {\textbf{Kernel: 100$\times$100}};
    \node at (3, 0.375+0.75-5-5) {\textbf{Bias: 100}};
    \node[text=white] at (3, 3-0.375-5-5) {\textbf{Fully Connected Layer}};
     \draw[->, >=stealth,line width=1.5pt] (3, 0-5-5) -- (3,-1-5-5);

    \draw[rounded corners,  draw=black] (0, -5-5-4) rectangle (6, -2-5-4);
    \draw[draw=black] (0, -2-0.75-5-4) -- (6, -2-0.75-5-4);
    \draw[draw=black] (0, -2-1.5-5-4) -- (6, -2-1.5-5-4);
    \draw[draw=black] (0, -2-2.25-5-4) -- (6, -2-2.25-5-4);
    \fill[rounded corners, darkred] (0, -2-2.25-5-4) rectangle (6, -5-5-4);
    \fill[rounded corners, darkblue] (0, -2-5-4) rectangle (6, -2-0.75-5-4);
    
    \node[text=white] at (3, 0.375-5-5-4) {\textbf{Activated by: Identity}};
    \node at (3, 0.375+0.75+0.75-5-5-4) {\textbf{Kernel: 100$\times$1}};
    \node at (3, 0.375+0.75-5-5-4) {\textbf{Bias: 1}};
    \node[text=white] at (3, 3-0.375-5-5-4) {\textbf{Output Layer}};
   
\end{tikzpicture}
\end{center}
    \caption{DeepKriging Model Visualization}
    \label{model:dk}
\end{figure}

This model, depicted by the graphic below, adds three resolutions of RBFs generated by the package \texttt{FRK}.  Bases 1-3 refer to the three RBF resolutions in the figure. No uniform basis function size exists since the number and dimension of RBFs depends on the spatial range of the data examined. Details about the neural network setting are shown in Figure \ref{model:dk}.

\subsection{SDCNN}
\begin{figure}
\begin{center}
\begin{tikzpicture}[scale = 0.8]


    \draw[rounded corners,  draw=black] (1, 5) rectangle (5, 4);
    \node at (3, 4.5) {\textbf{Basis-$i$}};
    \draw[->, >=stealth,line width=1.5pt] (3,4) -- (3,2+1);

    \draw[rounded corners,  draw=black] (-1, 0) rectangle (7, 2+1);
    \draw[draw=black] (-1, 0.75) -- (7, 0.75);
    \draw[draw=black] (-1, 1.5) -- (7, 1.5);
    \draw[draw=black] (-1, 2.25) -- (7, 2.25);
    \fill[rounded corners, rounded corners, darkred] (0-1, 0.75) rectangle (6+1, 0);
    \fill[rounded corners, darkblue] (0-1, 2+1) rectangle (6+1, 2.25);
    \node[text=white] at (3, 0.375) {\textbf{Activated by: ReLU}};
    \node at (3, 0.375+0.75+0.75) {\textbf{Kernel: $2\times2\times1\times128$}};
    \node at (3, 0.375+0.75) {\textbf{Bias: 128}};
    \node[text=white] at (3, 3-0.375) {\textbf{Input Layer(Conv-2D)}};
    \draw[->, >=stealth,line width=1.5pt] (3, 0) -- (3,-2);
    \draw[->, >=stealth,line width=1.5pt] (1.5, -1) -- (3,-1);
    \draw[rounded corners, draw=black](-4.5,-0.5) rectangle (1.5,-1.5);
    \node at (-1.5, -1) {\textbf{Flatten Layer}};

    \draw[rounded corners,  draw=black] (0, -5) rectangle (6, -2);
    \draw[draw=black] (0, -2-0.75) -- (6, -2-0.75);
    \draw[draw=black] (0, -2-1.5) -- (6, -2-1.5);
    \draw[draw=black] (0, -2-2.25) -- (6, -2-2.25);
    \fill[rounded corners, darkred] (0, -2-2.25) rectangle (6, -5);
    \fill[rounded corners, darkblue] (0, -2) rectangle (6, -2-0.75);
    
    \node[text=white] at (3, 0.375-5) {\textbf{Activated by: ReLU}};
    \node at (3, 0.375+0.75+0.75-5) {\textbf{Kernel: 128 $\times$100}};
    \node at (3, 0.375+0.75-5) {\textbf{Bias: 100}};
    \node[text=white] at (3, 3-0.375-5) {\textbf{Fully Connected Layer}};
    
    \draw[->, >=stealth,line width=1.5pt] (3, 0-5) -- (3,-2-5);
    \draw[->, >=stealth,line width=1.5pt] (7.5, -1-5) -- (3,-1-5);
    \draw[->, >=stealth,line width=1.5pt] (1.5, -1-5) -- (3,-1-5);
    \draw[rounded corners, draw=black](7.5,0-5) rectangle (10.5,-2-5);
    \fill[rounded corners, yellow!60](7.5,0-5) rectangle (10.5,-1-5);
    \node at (9, -0.5-5) {\textbf{Dropout}};
    \node at (9, -1.5-5) {\textbf{Rate = 0.1}};
    \draw[rounded corners, draw=black](-4.5,-0.5-5) rectangle (1.5,-1.5-5);
    \node at (-1.5, -1-5) {\textbf{Batch Normalization}};

    \draw[rounded corners,  draw=black] (0, -5-5) rectangle (6, -2-5);
    \draw[draw=black] (0, -2-0.75-5) -- (6, -2-0.75-5);
    \draw[draw=black] (0, -2-1.5-5) -- (6, -2-1.5-5);
    \draw[draw=black] (0, -2-2.25-5) -- (6, -2-2.25-5);
    \fill[rounded corners, darkred] (0, -2-2.25-5) rectangle (6, -5-5);
    \fill[rounded corners, darkblue] (0, -2-5) rectangle (6, -2-0.75-5);
    
    \node[text=white] at (3, 0.375-5-5) {\textbf{Activated by: ReLU}};
    \node at (3, 0.375+0.75+0.75-5-5) {\textbf{Kernel: 100$\times$100}};
    \node at (3, 0.375+0.75-5-5) {\textbf{Bias: 100}};
    \node[text=white] at (3, 3-0.375-5-5) {\textbf{Fully Connected Layer}};
    \draw[->, >=stealth,line width=1.5pt] (3, 0-5-5) -- (3,-1-5-5);
    
    \draw[->, >=stealth,line width=1.5pt] (3, 0-5-5) -- (3,-2-5-5);
    \draw[->, >=stealth,line width=1.5pt] (7.5, -1-5-5) -- (3,-1-5-5);
    \draw[->, >=stealth,line width=1.5pt] (1.5, -1-5-5) -- (3,-1-5-5);
    \draw[rounded corners, draw=black](7.5,0-5-5) rectangle (10.5,-2-5-5);
    \fill[rounded corners, yellow!60](7.5,0-5-5) rectangle (10.5,-1-5-5);
    \node at (9, -0.5-5-5) {\textbf{Dropout}};
    \node at (9, -1.5-5-5) {\textbf{Rate = 0.1}};
    \draw[rounded corners, draw=black](-4.5,-0.5-5-5) rectangle (1.5,-1.5-5-5);
    \node at (-1.5, -1-5-5) {\textbf{Batch Normalization}};

    \draw[rounded corners,  draw=black] (0, -5-5-5) rectangle (6, -2-5-5);
    \draw[draw=black] (0, -2-0.75-5-5) -- (6, -2-0.75-5-5);
    \draw[draw=black] (0, -2-1.5-5-5) -- (6, -2-1.5-5-5);
    \draw[draw=black] (0, -2-2.25-5-5) -- (6, -2-2.25-5-5);
    \fill[rounded corners, darkred] (0, -2-2.25-5-5) rectangle (6, -5-5-5);
    \fill[rounded corners, darkblue] (0, -2-5-5) rectangle (6, -2-0.75-5-5);
    
    \node[text=white] at (3, 0.375-5-5-5) {\textbf{Activated by: ReLU}};
    \node at (3, 0.375+0.75+0.75-5-5-5) {\textbf{Kernel: 100$\times$100}};
    \node at (3, 0.375+0.75-5-5-5) {\textbf{Bias: 100}};
    \node[text=white] at (3, 3-0.375-5-5-5) {\textbf{Fully Connected Layer}};

\node at (3, 3-0.375-5-5-5-5+0.25) {\textbf{Convatenate Layer}};
    \draw[rounded corners, draw=black](0,3-0.375-5-5-5-5-1+0.5) rectangle (6, 3-0.375-5-5-5-5+1);
    \draw[->, >=stealth,line width=1.5pt] (3, 3-0.375-5-5-5-5+2.5-0.125) -- (3, 3-0.375-5-5-5-5+1);
    \draw[->, >=stealth,line width=1.5pt] (3, 3-0.375-5-5-5-5+2.5) -- (3, 3-0.375-5-5-5-5+1);
\end{tikzpicture}
\end{center}
\caption{SDCNN Model Visualization of Convolutional Part.}
\label{model:SDCNN1}
\end{figure}
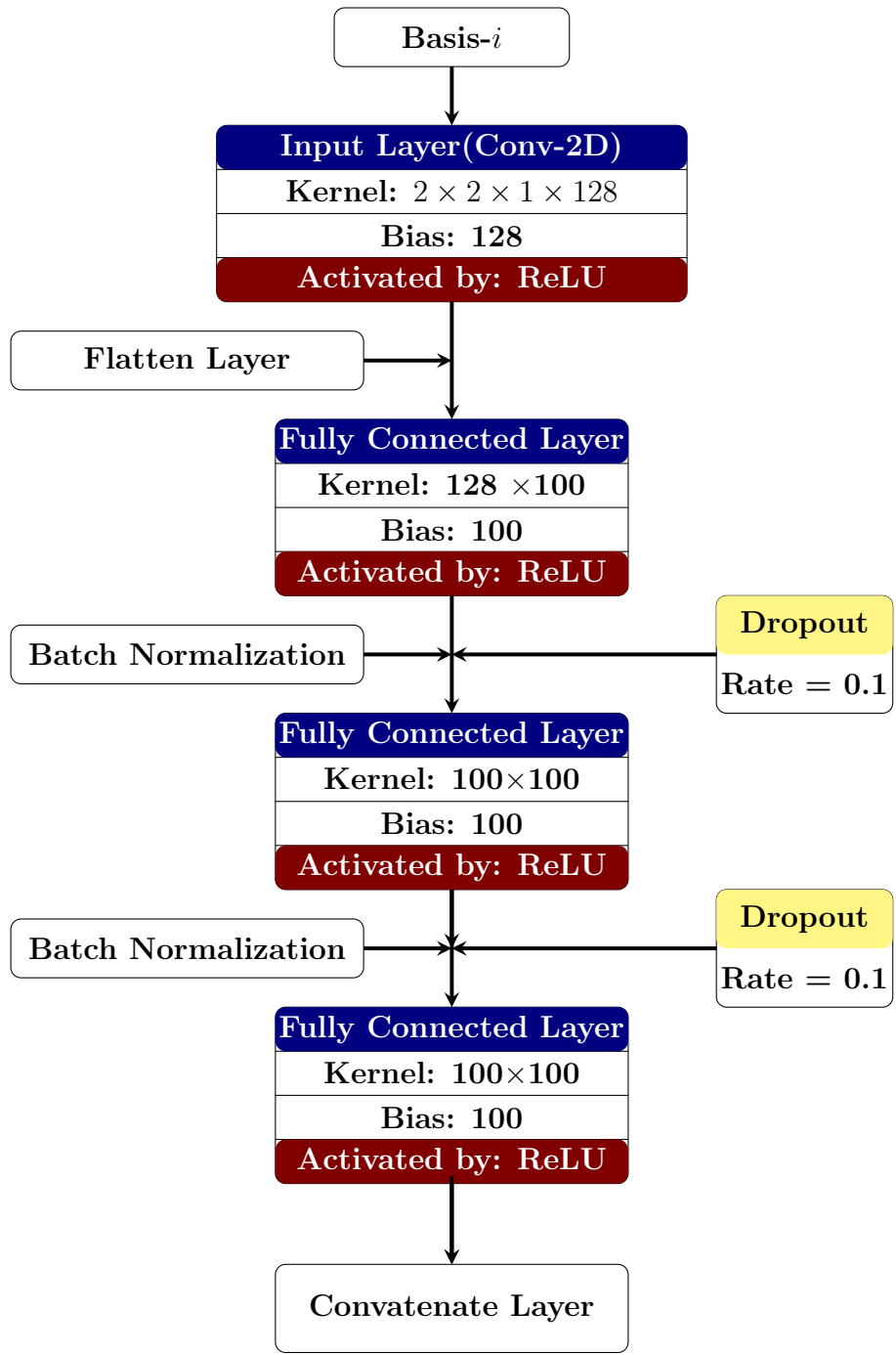

This model first convolves each of the three RBF resolutions, as shown in Figure \ref{model:SDCNN1}. These are subsequently concatenated via longitude and latitude. Each resolution uses a similar structure; hence, only one resolution is displayed in the figure. Here, the resolution $i$ RBFs have $m_i$ rows and $n_i$ columns. The three resolutions and coordinates are calculated individually before concatenation.

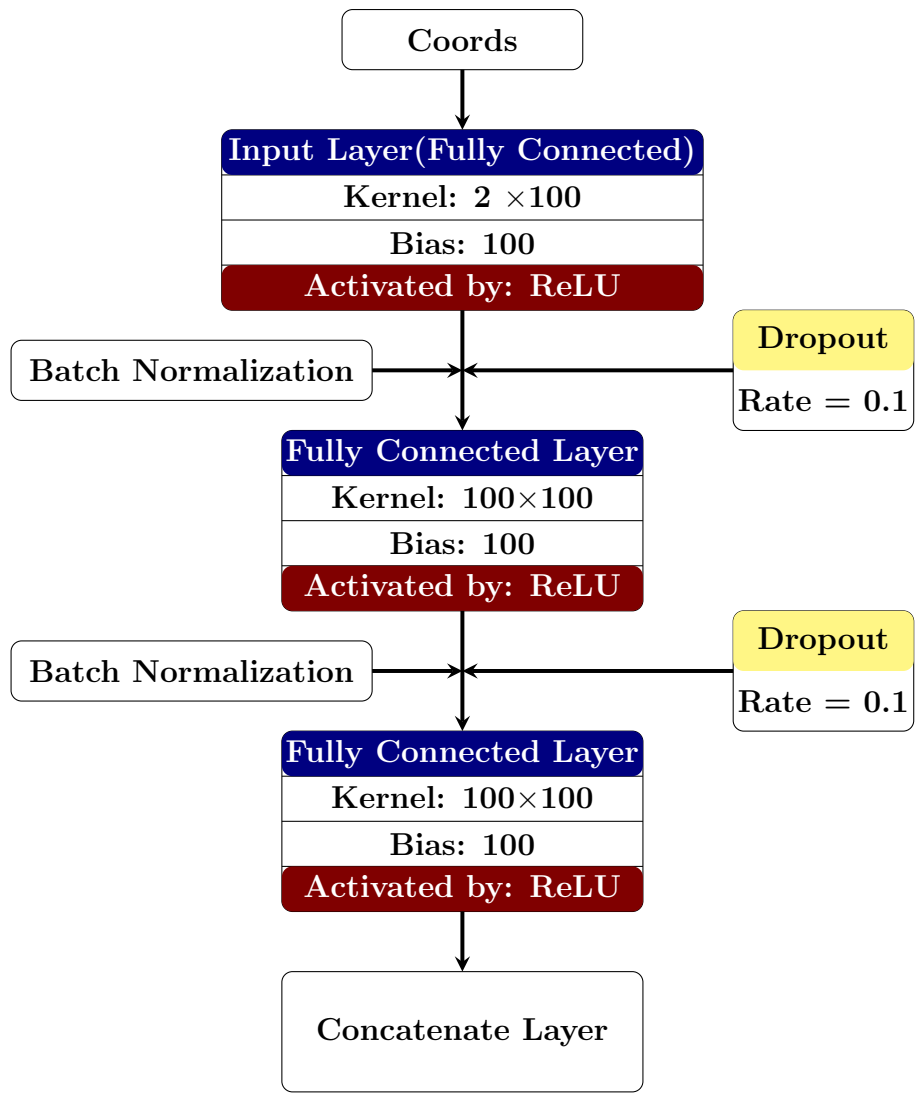
\begin{figure}
    \begin{center}
\begin{tikzpicture}[scale = 0.8]

    \draw[rounded corners,  draw=black] (1, 5) rectangle (5, 4);
    \node at (3, 3.5+1) {\textbf{Coords}};
    \draw[->, >=stealth,line width=1.5pt] (3, 3+1) -- (3,2+1);
    
    \draw[rounded corners,  draw=black] (-1, 0) rectangle (7, 2+1);
    \draw[draw=black] (-1, 0.75) -- (7, 0.75);
    \draw[draw=black] (-1, 1.5) -- (7, 1.5);
    \draw[draw=black] (-1, 2.25) -- (7, 2.25);
    \fill[rounded corners, darkred] (0-1, 0.75) rectangle (6+1, 0);
    \fill[rounded corners, darkblue] (0-1, 2+1) rectangle (6+1, 2.25);
    \node[text=white] at (3, 0.375) {\textbf{Activated by: ReLU}};
    \node at (3, 0.375+0.75+0.75) {\textbf{Kernel: 2 $\times$100}};
    \node at (3, 0.375+0.75) {\textbf{Bias: 100}};
    \node[text=white] at (3, 3-0.375) {\textbf{Input Layer(Fully Connected)}};
    \draw[->, >=stealth,line width=1.5pt] (3, 0) -- (3,-2);
    \draw[->, >=stealth,line width=1.5pt] (7.5, -1) -- (3,-1);
    \draw[->, >=stealth,line width=1.5pt] (1.5, -1) -- (3,-1);
    \draw[rounded corners, draw=black](7.5,0) rectangle (10.5,-2);
    \fill[rounded corners, yellow!60](7.5,0) rectangle (10.5,-1);
    \node at (9, -0.5) {\textbf{Dropout}};
    \node at (9, -1.5) {\textbf{Rate = 0.1}};
    \draw[rounded corners, draw=black](-4.5,-0.5) rectangle (1.5,-1.5);
    \node at (-1.5, -1) {\textbf{Batch Normalization}};

    \draw[rounded corners,  draw=black] (0, -5) rectangle (6, -2);
    \draw[draw=black] (0, -2-0.75) -- (6, -2-0.75);
    \draw[draw=black] (0, -2-1.5) -- (6, -2-1.5);
    \draw[draw=black] (0, -2-2.25) -- (6, -2-2.25);
    \fill[rounded corners, darkred] (0, -2-2.25) rectangle (6, -5);
    \fill[rounded corners, darkblue] (0, -2) rectangle (6, -2-0.75);
    
    \node[text=white] at (3, 0.375-5) {\textbf{Activated by: ReLU}};
    \node at (3, 0.375+0.75+0.75-5) {\textbf{Kernel: 100$\times$100}};
    \node at (3, 0.375+0.75-5) {\textbf{Bias: 100}};
    \node[text=white] at (3, 3-0.375-5) {\textbf{Fully Connected Layer}};
    
    \draw[->, >=stealth,line width=1.5pt] (3, 0-5) -- (3,-2-5);
    \draw[->, >=stealth,line width=1.5pt] (7.5, -1-5) -- (3,-1-5);
    \draw[->, >=stealth,line width=1.5pt] (1.5, -1-5) -- (3,-1-5);
    \draw[rounded corners, draw=black](7.5,0-5) rectangle (10.5,-2-5);
    \fill[rounded corners, yellow!60](7.5,0-5) rectangle (10.5,-1-5);
    \node at (9, -0.5-5) {\textbf{Dropout}};
    \node at (9, -1.5-5) {\textbf{Rate = 0.1}};
    \draw[rounded corners, draw=black](-4.5,-0.5-5) rectangle (1.5,-1.5-5);
    \node at (-1.5, -1-5) {\textbf{Batch Normalization}};

    \draw[rounded corners,  draw=black] (0, -5-5) rectangle (6, -2-5);
    \draw[draw=black] (0, -2-0.75-5) -- (6, -2-0.75-5);
    \draw[draw=black] (0, -2-1.5-5) -- (6, -2-1.5-5);
    \draw[draw=black] (0, -2-2.25-5) -- (6, -2-2.25-5);
    \fill[rounded corners, darkred] (0, -2-2.25-5) rectangle (6, -5-5);
    \fill[rounded corners, darkblue] (0, -2-5) rectangle (6, -2-0.75-5);
    
    \node[text=white] at (3, 0.375-5-5) {\textbf{Activated by: ReLU}};
    \node at (3, 0.375+0.75+0.75-5-5) {\textbf{Kernel: 100$\times$100}};
    \node at (3, 0.375+0.75-5-5) {\textbf{Bias: 100}};
    \node[text=white] at (3, 3-0.375-5-5) {\textbf{Fully Connected Layer}};
    \draw[->, >=stealth,line width=1.5pt] (3, 0-5-5) -- (3,-1-5-5);
    
    \node at (3, 3-5-5-5) {\textbf{Concatenate Layer}};
    \draw[rounded corners, draw=black](0,-1-5-5) rectangle (6,-3-5-5);
    
\end{tikzpicture}
\end{center}
\caption{SDCNN Model Visualization of Covariates Part.}
\label{model:SDCNN2}
\end{figure}

\begin{figure}
    \begin{center}
\begin{tikzpicture}[scale = 0.8]

    \draw[rounded corners,  draw=black] (-6, 5+1+1+1) rectangle (-3, 4+1+1);
    \draw[rounded corners,  draw=black] (12, 5+1+1+1) rectangle (9, 4+1+1);
    \draw[rounded corners,  draw=black] (-1, 5+1+1+1) rectangle (2, 4+1+1);
    \draw[rounded corners,  draw=black] (4, 5+1+1+1) rectangle (7, 4+1+1);
    \node at (-4.5, 4.5+1+1+1-0.25) {\textbf{Coords}};
    \node at (-4.5, 4.5+1+1+0.25) {\textbf{Submodel}};
    \node at (-4.5+5, 4.5+1+1+1-0.25) {\textbf{Basis-1}};
    \node at (-4.5+5, 4.5+1+1+0.25) {\textbf{Submodel}};
    \node at (-4.5+5+5, 4.5+1+1+1-0.25) {\textbf{Basis-2}};
    \node at (-4.5+5+5, 4.5+1+1+0.25) {\textbf{Submodel}};
    \node at (-4.5+5+5+5, 4.5+1+1+1-0.25) {\textbf{Basis-3}};
    \node at (-4.5+5+5+5, 4.5+1+1+0.25) {\textbf{Submodel}};
    
    \draw[->, >=stealth,line width=1.5pt] (-4.5,4+1+1) -- (3,2+1+1);
    \draw[->, >=stealth,line width=1.5pt] (-4.5+5,4+1+1) -- (3,2+1+1);
    \draw[->, >=stealth,line width=1.5pt] (-4.5+5+5,4+1+1) -- (3,2+1+1);
    \draw[->, >=stealth,line width=1.5pt] (-4.5+5+5+5,4+1+1) -- (3,2+1+1);

    \node at (3,2+1+1-0.5) {\textbf{Concatenate Layer}};
    \draw[rounded corners,  draw=black] (0,2+1+1) rectangle (6,2+1);
    \draw[->, >=stealth,line width=1.5pt] (3,2+1) -- (3,2);

    \draw[rounded corners,  draw=black] (-1, 0-1) rectangle (7, 2+1-1);
    \draw[draw=black] (-1, 0.75-1) -- (7, 0.75-1);
    \draw[draw=black] (-1, 1.5-1) -- (7, 1.5-1);
    \draw[draw=black] (-1, 2.25-1) -- (7, 2.25-1);
    \fill[rounded corners, darkred] (0-1, 0.75-1) rectangle (6+1, 0-1);
    \fill[rounded corners, darkblue] (0-1, 2+1-1) rectangle (6+1, 2.25-1);
    \node[text=white] at (3, 0.375-1) {\textbf{Activated by: Identity}};
    \node at (3, 0.375+0.75+0.75-1) {\textbf{Kernel: 400 $\times$1}};
    \node at (3, 0.375+0.75-1) {\textbf{Bias: 1}};
    \node[text=white] at (3, 3-0.375-1) {\textbf{Output Layer}};

\end{tikzpicture}
\end{center}
\caption{SDCNN Model Visualization of Concatenating and Output Layers}
\label{model:SDCNN3}
\end{figure}
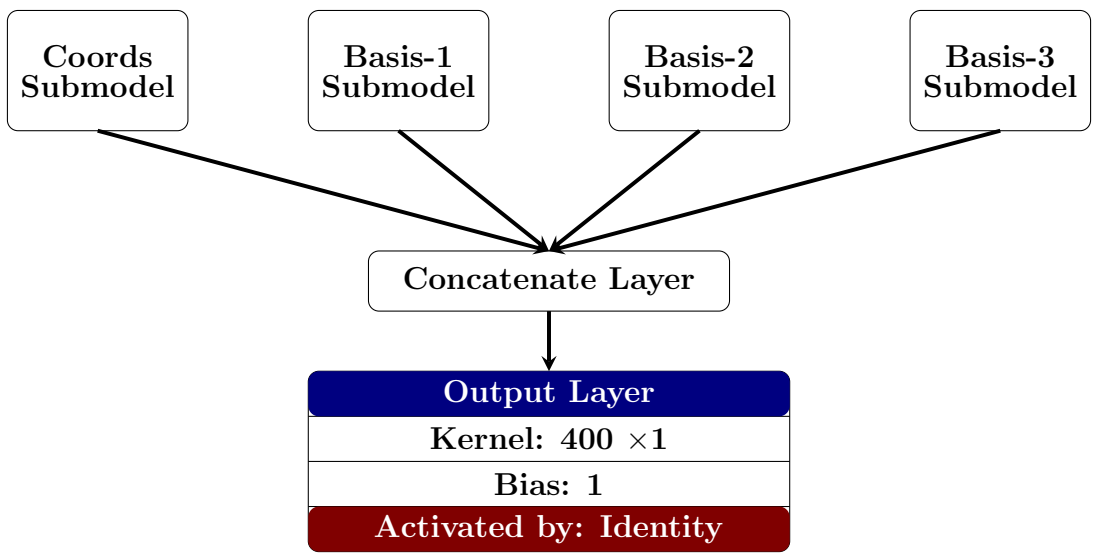

After the three RBF resolutions, a model similar in structure to a baseline DNN is used for the longitudes and latitudes, concatenating these together. This procedure is graphically depicted in Figure \ref{model:SDCNN2}.
With all three RBF resolutions and coordinates dealt with, our next step concatenates these and feeds the result through an output layer. Figure \ref{model:SDCNN3} depicts the procedure.

\end{document}